\documentclass[11pt,a4paper]{article}

\usepackage{tabto}
\usepackage{tgpagella,eulervm}
\usepackage[bottom,flushmargin,hang,multiple]{footmisc}
\usepackage[onehalfspacing]{setspace}
\usepackage[left=2cm, right=2cm, top=3cm, bottom=2cm]{geometry}
\usepackage[font=small,labelfont=bf]{caption}
\usepackage{subcaption}
\usepackage{color,soul}

\usepackage{lastpage}
\usepackage{changes}
\usepackage{natbib}
\usepackage{hyperref}
\usepackage{amsmath}
\usepackage{amssymb}
\usepackage{mathtools}
\usepackage{eurosym}
\usepackage{booktabs}
\usepackage{appendix}
\usepackage{xcolor}
\usepackage{enumitem}
\usepackage{rotating}
\usepackage{graphicx}
\usepackage{relsize}
\usepackage{courier}

\usepackage{zref-totpages}

    \makeatletter
\def\@fnsymbol#1{\ensuremath{\ifcase#1\or \mathsection\or \mathparagraph\or \|\or **\or \dagger\dagger
   \or \ddagger\ddagger \else\@ctrerr\fi}}
    \makeatother

\title{\huge \textbf{Introducing advanced hybrid coupling: Non-discriminatory coalescence of flow-based and net transfer capacity calculation regions}}
\author{...}
\date{\today{}}

\author{
David Sch\"onheit \thanks{david.schoenheit@50hertz.com, 50Hertz Transmission GmbH}
\and
Ivan Marjanovi\'c \thanks{imarjanovic@e-bridge.com, E-Bridge Consulting GmbH} }
\date{\small \textbf{Working paper}: Version of \today{}
}

\usepackage{fancyhdr}
\definecolor{MK}{rgb}{0,0,1}

\begin{document}

\clearpage
\maketitle

\thispagestyle{empty}

{\small \textbf{Abstract}: Flow-based market coupling is substantially altering the computation of cross-zonal capacities for the trade of electricity in the vast majority of European markets. The main benefit of the flow-based method is improved accuracy by better representing the impact of cross-zonal trade on the power flows in transmission grids. Some borders, adjacent to flow-based capacity regions, are represented through net transfer capacities during market coupling. Under the current standard hybrid coupling, the utilization of grid elements in the flow-based regions due to the predicted trade across such borders is not available for trades between flow-based zones. The flow-based representation is not limited to the given capacity calculation region, but can be extended to also model the impact of trade with other regions. This so-called advanced hybrid coupling replaces the priority inherently given to trade across net transfer capacity-coupled borders by introducing virtual bidding zones. These map the effect of non-flow-based borders on line capacities in the flow-based regions, enabling the market coupling optimization to prioritize trade between flow-based bidding zones and trade across non-flow-based borders. This paper explains the mechanism of advanced hybrid coupling and how it is modeled mathematically. Based on a test network, a case study shows to what extent and why advanced hybrid coupling leads to welfare gains during market coupling and lower congestion management costs in the flow-based region.


}

\section*{Nomenclature}
\label{app:nom}

{\footnotesize
\begin{spacing}{1}
\bigskip
\noindent \textbf{Indices and sets}:
\begin{itemize}[itemsep=0.1em, parsep=0pt, label={--}]
\item $Z$: Physical bidding zones $z \in Z$
\begin{itemize}[itemsep=0.1em, parsep=0pt, label={--}]
\item $Z^\mathrm{FB}$ \tabto{1.75cm} Flow-based market coupling zones, $Z^{FB} \subset Z$
\end{itemize}
\item $Z_{\mathrm{AHC}}^{\mathrm{FB}}$: Virtual/physical flow-based bidding zones in AHC $z \in Z_{\mathrm{AHC}}^{\mathrm{FB}}$
\begin{itemize}[itemsep=0.1em, parsep=0pt, label={--}]
\item $Z_{\mathrm{AHC, phys}}^{\mathrm{FB}}$ \tabto{1.75cm} Physical flow-based bidding zones $Z^{FB} = Z_{\mathrm{AHC, phys}}^{\mathrm{FB}} \subset Z_{\mathrm{AHC}}^{\mathrm{FB}}$
\item $Z_{\mathrm{AHC, virt}}^{\mathrm{FB}}$ \tabto{1.75cm} Virtual flow-based bidding zones, $Z_{\mathrm{AHC, virt}}^{\mathrm{FB}} \subset Z_{\mathrm{AHC}}^{\mathrm{FB}}$
\end{itemize}
\item $L$: AC lines $l \in L$
\begin{itemize}[itemsep=0.1em, parsep=0pt, label={--}]
\item $J$ \tabto{1.75cm} Critical network elements under contingencies CNECs $j \in J$
\item $L^\mathrm{c}$ \tabto{1.75cm} Set of lines $L$ extended by lines under contingencies 
\end{itemize}
\item $N$: Grid nodes $n \in N$
\begin{itemize}[itemsep=0.1em, parsep=0pt, label={--}]
\item $V$ \tabto{1.75cm} Slack node $v \in V \subset N$
\item $N^\mathrm{FB}$ \tabto{1.75cm} Flow-based market coupling nodes, $N_{FB} \subset N$
\item $N^\mathrm{nFB}$ \tabto{1.75cm} Non-flow-based market coupling nodes, $N_{nFB} \subset N$
\end{itemize}
\item $P$: Power plants $p \in P$
\begin{itemize}[itemsep=0.1em, parsep=0pt, label={--}]
\item $P^\mathrm{RD}$ \tabto{1.75cm} Redispatch power plants $p \in P^\mathrm{RD} \subset P$
\end{itemize}
\item $T$: Timesteps/hours $t \in T := \{1, 2, ..., 8760\}$
\end{itemize}

\bigskip
\noindent \textbf{Parameters}:
\begin{itemize}[itemsep=0.1em, parsep=0pt, label={--}]
\item Power plant characteristics: 
\begin{itemize}[itemsep=0.1em, parsep=0pt, label={--}]
\item $g_{p}^\mathrm{max}$ \tabto{1.75cm} Maximum power output (installed capacity in MW) $g_{p}^\mathrm{max} \in \mathbb{R}_{> 0}$
\item $\mathbf{G}$ / $\mathbf{G^\mathrm{AHC}}$ \tabto{1.75cm} Generation shift key matrix $\mathbf{G} \in [-1,1]^{|N| \times |Z|}$ / $\mathbf{G^\mathrm{AHC}} \in [-1,1]^{|N| \times |Z_{\mathrm{AHC}}^{\mathrm{FB}}|}$
\end{itemize}
\item Costs and penalties (all in EUR/MWh):
\begin{itemize}[itemsep=0.1em, parsep=0pt, label={--}]
\item $c_{t,p}^\mathrm{var}$ \tabto{1.75cm} Variable (generation) costs $c_{t,p}^\mathrm{var} \in \mathbb{R}_{\geq 0}$ 
\item $c^{\mathrm{RD, pos}}_{t,p}$ \tabto{1.75cm} Positive redispatch penalty $c^{\mathrm{RD, pos}}_{t,p} \in \mathbb{R}_{\geq 0}$ 
\item $c^{\mathrm{RD, neg}}_{t,p}$ \tabto{1.75cm} Negative redispatch penalty $c^{\mathrm{RD, neg}}_{t,p} \in \mathbb{R}_{\geq 0}$ 
\item $c^{\mathrm{curt}}$ \tabto{1.75cm} Penalty for curtailed renewable energy $c^{\mathrm{curt}} \in \mathbb{R}_{\geq 0}$ 
\end{itemize}
\item Power line characteristics:
\begin{itemize}[itemsep=0.1em, parsep=0pt, label={--}]
\item $\mathbf{P^\mathrm{\mathbf{N}}}$ \tabto{1.75cm} Nodal PTDF matrix $\mathbf{P^\mathrm{\mathbf{N}}} \in [-1,1]^{|L| \times |N|}$
\item $\mathbf{P^\mathrm{\mathbf{Z}}}$ / $\mathbf{P_{\mathrm{AHC}}^\mathrm{\mathbf{Z}}}$ \tabto{1.75cm} Zodal PTDF matrix $\mathbf{P^\mathrm{\mathbf{Z}}} \in [-1,1]^{|L| \times |Z|}$ / $\mathbf{P_{\mathrm{AHC}}^\mathrm{\mathbf{Z}}} \in [-1,1]^{|L| \times |Z_{\mathrm{AHC}}^{\mathrm{FB}}|}$
\item $fmax_{l}$ \tabto{1.75cm} Capacity limit of AC line (MW) $fmax_{l} \in \mathbb{R}_{> 0}$ 
\item $frm_{j}$ \tabto{1.75cm} Flow reliability margins $frm_{j} \in \mathbb{R}_{\geq 0}$ 
\end{itemize}
\item Exogenous time series (MW):
\begin{itemize}[itemsep=0.1em, parsep=0pt, label={--}]
\item $ren_{t,n}$ \tabto{1.75cm} Available node-specific renewable generation
\item $ren_{t,n}^\mathrm{D-2}$ \tabto{1.75cm} Available node-specific renewable generation (D-2 forecast)
\item $d_{t,n}$ \tabto{1.75cm} Demand per node $d_{t,n} \in \mathbb{R}_{\geq 0}$
\end{itemize}
\item Trade:
\begin{itemize}[itemsep=0.1em, parsep=0pt, label={--}]
\item $ntc_{t,z}$ \tabto{1.75cm} Max. export capacity from zone $z \notin Z_\mathrm{FB}$ $ntc_{t,z,x} \in \mathbb{R}_{\geq 0}$
\end{itemize}
\end{itemize}

\bigskip
\noindent \textbf{Variables} (Many exist for D-2, D-1 and D-0, indicated superscripts in the model notations):
\begin{itemize}[itemsep=0.1em, parsep=0pt, label={--}]
\item Cost variables:
\begin{itemize}[itemsep=0.1em, parsep=0pt, label={--}]
\item $TC$ \tabto{1.75cm} Total costs $TC \in \mathbb{R}_{\geq 0}$
\item $CG_{t,z}$ \tabto{1.75cm} Generation costs $CG_{t,z} \in \mathbb{R}_{\geq 0}$
\item $CC_{t,z}$ \tabto{1.75cm} Curtailment costs $CC_{t,z} \in \mathbb{R}_{\geq 0}$
\end{itemize}
\item Market model variables:
\begin{itemize}[itemsep=0.1em, parsep=0pt, label={--}]
\item $G_{t,p}$ \tabto{1.75cm} Power plant generation $G_{t,p} \in \mathbb{R}_{\geq 0}$
\item $CURT_{t,n}$ \tabto{1.75cm} Curtailment $CURT_{t,n} \in \mathbb{R}_{\geq 0}$
\end{itemize}
\item Redispatch model variables:
\begin{itemize}[itemsep=0.1em, parsep=0pt, label={--}]
\item $RD_{t,p}^\mathrm{pos}$ \tabto{1.75cm} Positive redispatch $RD_{t,p}^\mathrm{pos} \in \mathbb{R}_{\geq 0}$
\item $RD_{t,p}^\mathrm{neg}$ \tabto{1.75cm} Negative redispatch $RD_{t,p}^\mathrm{neg} \in \mathbb{R}_{\geq 0}$
\end{itemize}
\item Grid variables:
\begin{itemize}[itemsep=0.1em, parsep=0pt, label={--}]
\item $INJ_{t,n}$ \tabto{1.75cm} Net injection $INJ_{t,n} \in \mathbb{R}$
\item $F_{t,l}$ \tabto{1.75cm} Power flow on AC line $F_{t,l} \in \mathbb{R}$
\end{itemize}
\item Trade variables:
\begin{itemize}[itemsep=0.1em, parsep=0pt, label={--}]
\item $EX_{t,z}$ \tabto{1.75cm} Export from or import to (+/- values) non-FB zone $z \notin Z_\mathrm{FB}$ $EX_{t,z} \in \mathbb{R}$
\item $NP_{t,z}$ \tabto{1.75cm} Net position in flow-based market coupling area $NP_{t,z} \in \mathbb{R}$
\end{itemize}
\end{itemize}

\bigskip
\noindent \textbf{Mapping}:
\begin{itemize}[itemsep=0.1em, parsep=0pt, label={--}]
\item $mn(z)$ \tabto{1.75cm} $z \in Z \Rightarrow n \in N$ \tabto{6cm} All nodes in a zone
\item $mz(z)$ \tabto{1.75cm} $z \in Z \Rightarrow z \in Z$ \tabto{6cm} All zones connected to a zone through an NTC border
\item $mvbz(z)$ \tabto{1.75cm} $z \in Z_{\mathrm{AHC, virt}}^{\mathrm{FB}} \Rightarrow z \not\in Z^{\mathrm{FB}}$ \tabto{6cm} All non-FB zones connected to an AHC VBZ
\item $mp(n)$ \tabto{1.75cm} $n \in N \Rightarrow p \in P$ \tabto{6cm} All power plants at a node
\item $mp(z)$ \tabto{1.75cm} $z \in Z \Rightarrow p \in P$ \tabto{6cm} All power plants in a zone
\item $mprd(n)$ \tabto{1.75cm} $n \in N \Rightarrow p \in P^{\mathrm{RD}}$ \tabto{6cm} All redispatch power plants at a node
\item $mprd(z)$ \tabto{1.75cm} $z \in Z \Rightarrow p \in P^{\mathrm{RD}}$ \tabto{6cm} All redispatch power plants in a zone
\end{itemize}

\end{spacing}
}

\newpage
\part{Motivation and theoretical background of advanced hybrid coupling}
\section{Introduction and objectives}
\label{sec:intro}
\begin{figure}[!h]
    \centering  \includegraphics[width=1\linewidth]{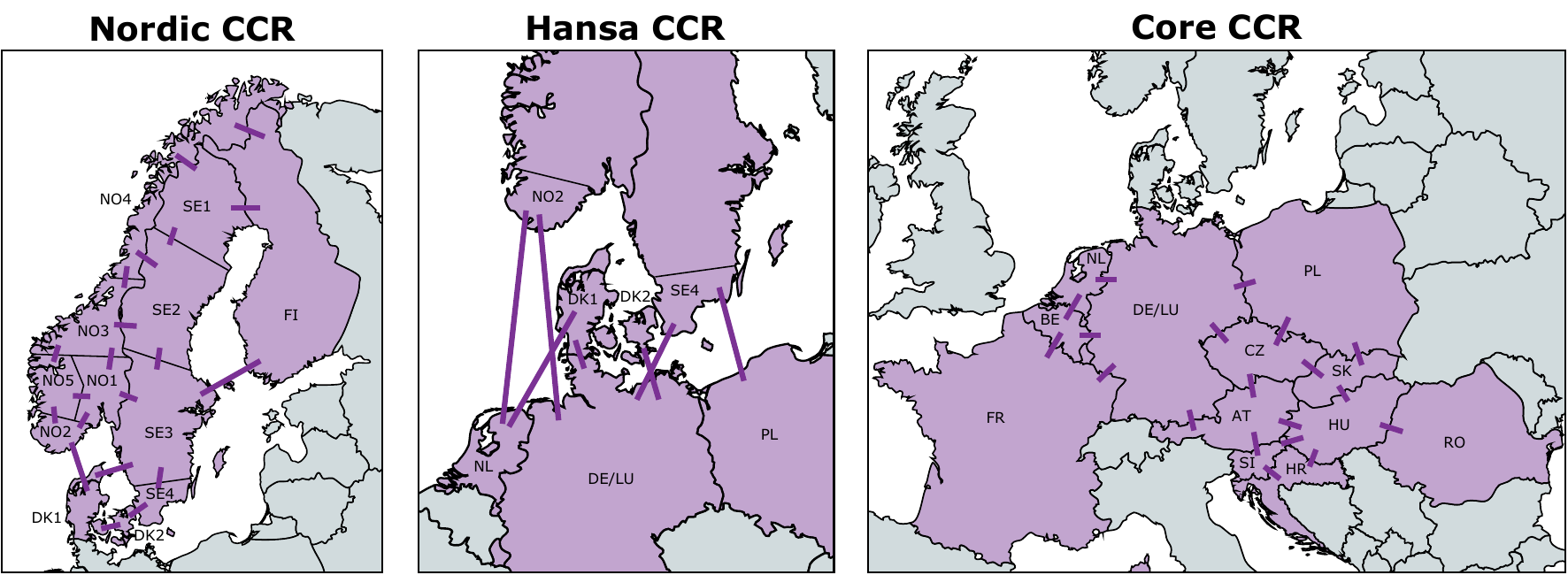}
    \caption{Capacity calculation regions Nordic, Hansa and Core. Bidding zones and CCRs based on \cite{ccrs} and maps made with \cite{MapCharts}. Of the Norwegian and Swedish bidding zones, only borders including NO2 and SE4 are part of the Hansa CCR.}
    \label{fig:ccr}
\end{figure}

Coordinated capacity calculation in Europe is performed on a regional level, in the scope of capacity calculation regions (CCRs). In total there are ten CCRs in Europe. For Core CCR and Nordic CCR, depicted on Fig. \ref{fig:ccr}, as two regions characterized by a meshed interconnected grid and highly interdependent cross-zonal capacities on individual borders, it has been decided to apply the flow-based (FB) method for capacity calculation (CC) and allocation. The flow-based method is the most efficient CC method for meshed networks, as it has several advantages over the coordinated net transfer capacity (CNTC) method \cite{acer_2016}:

\begin{enumerate}
    \item \textbf{It enables optimal allocation of cross-zonal capacity}:  Capacity is allocated in an optimized way (e.g. maximizing social welfare) during the market coupling (MC). In the CNTC method, capacity on individual borders is split “ex-ante” (before MC) as the interdependencies between individual borders cannot be modeled. An ”ex-ante” split is prone to forecasting errors and particularly inefficient in meshed networks.
   \item \textbf{It offers better transparency as the physical congestions can be easily identified}: The flow-based method models the transmission capacities as a set of constraints that represent physical power flow limits in the transmission grid. Each constraint corresponds to one particular grid element (critical network elements associated with a contingency = CNEC), hence enabling the identification of congested elements that are limiting cross-zonal trade.
\end{enumerate}

Besides modeling the interdependencies between the cross-zonal capacities on the borders within the given CCR, the FB method also enables to represent the impact of the borders to adjacent CCRs by introducing “virtual bidding zones” (VBZ) on the nodes of corresponding interconnectors. This concept is called advanced hybrid coupling (AHC) and through its usage it is possible to exploit the mentioned benefits of the FB method also on the borders to adjacent CCRs. AHC is the target solution for modeling the Hansa CCR borders, which are located between Core and Nordic CCR. "The expectation is that by ensuring a non-discriminatory competition for the scarce CNEC capacity, AHC will lead to an increase in socio-economic welfare and improved operational gird security at the same time" as stated by the explanatory document to the second amendment of the Day-Ahead Capacity Calculation Methodology of the Core Capacity Calculation Region \citep{entsoe_2}.

\begin{figure}[!h]
    \centering
\includegraphics[width=1\textwidth]{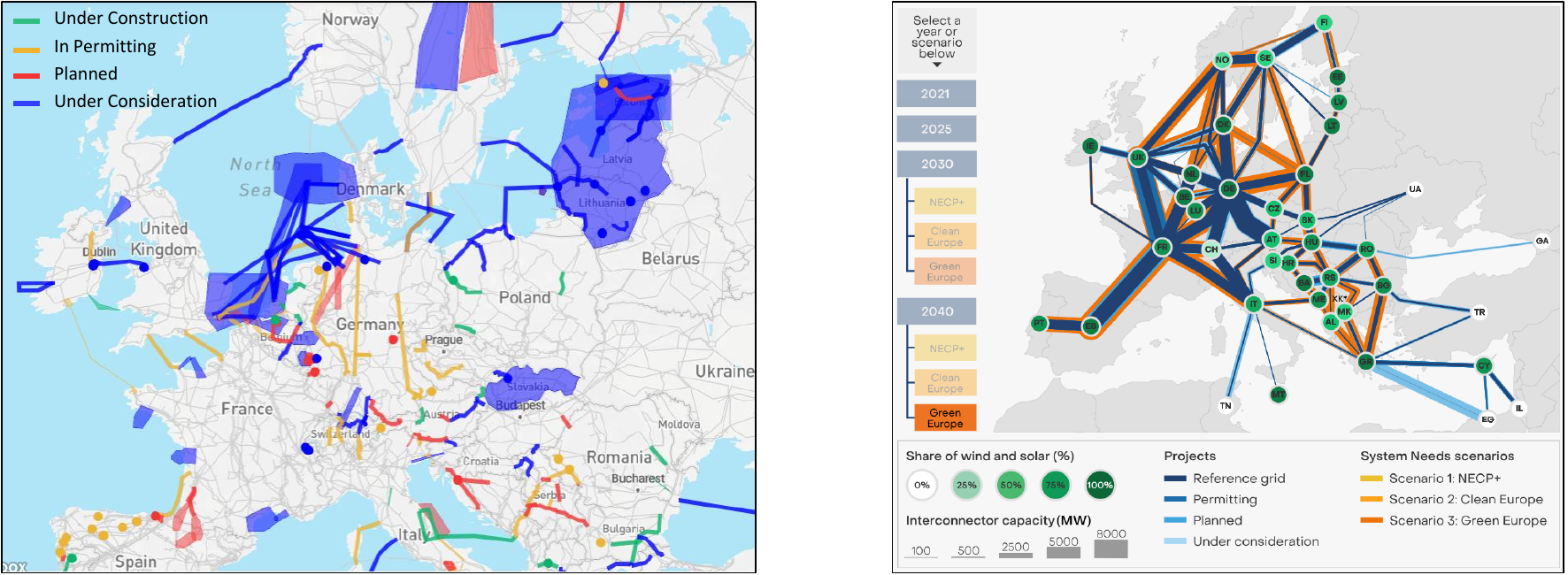}
    \caption{Left: Transmission system projects in Europe based on \cite{tyndp}. Right: Expected and required interconnection capacity for the energy transition scenario "2040 Green Europe" \citep{ember}.}
    \label{fig:tyndp}
\end{figure}

These benefits are estimated to be significant already today, as there is a strong coupling between the mentioned CCRs due to high interconnection capacity. Furthermore, a considerable increase of transmission capacity is expected in the near future due to large number of planned expansion projects, as depicted in Fig. \ref{fig:tyndp}. Further interconnection projects are required to reach envisioned energy transitions.

The introduction of AHC in the day-ahead market is planned in both Nordic CCR and Core CCR before the end of 2025 \citep{entsoe_3, nordic_golive_update}, which will have a considerable impact on almost all stakeholders in European electricity markets. However, the concept of AHC is mostly new to the stakeholders and it is still rarely described in the scientific and technical literature.

The objective of this paper is to provide a theoretical introduction to advanced hybrid coupling, starting with a technical description and assessment of its benefits compared to the currently applied standard hybrid coupling (SHC). Furthermore, the paper includes a case study based on a test network, which provides exemplary quantitative assessments on the impact of AHC regarding the most relevant aspects of the FB capacity calculation and allocation.

\section{Hybrid coupling: Standard vs advanced}
\label{sec:shc_ahc}
\subsection{Background and general provisions}
\label{shc_ahc:sub1}
During the implementation of the FB method in the Central Western Europe (CWE) region, the concept of “FB/ATC Hybrid Price Coupling” was introduced in 2011 (see \citealp{cwe_enhanced_fbmc_report_2011}). It was addressing the challenge of how to integrate the flow-based method into the capacity allocation process, e.g. how to combine FB and ATC constraints coming from different CCRs. The main difficulty for the hybrid coupling approach was to “fairly take the influence of one model over another.” Back then it was distinguished between the “rough” and “advanced” FB/ATC hybrid price coupling:
\begin{itemize}[itemsep=0.1em, parsep=0pt, label={--}]
\item In the “rough” method realized ATC transactions are not taken into account in the physical margins of critical network elements in the FB model. Therefore, in order to ensure operational security, the most constraining scenario of allocated ATC has to be reserved in the physical margin of CNEs in the FB model.
\item “Advanced” method takes into account the impact of realized ATC transactions on the CNE physical margins during allocation. Thus, no priority is given to ATC transaction compared to FB transactions and the usage of physical margins becomes a market decision with no reservation in advance. 
\end{itemize}

\begin{figure}[h!]
\includegraphics[width=0.95\textwidth]{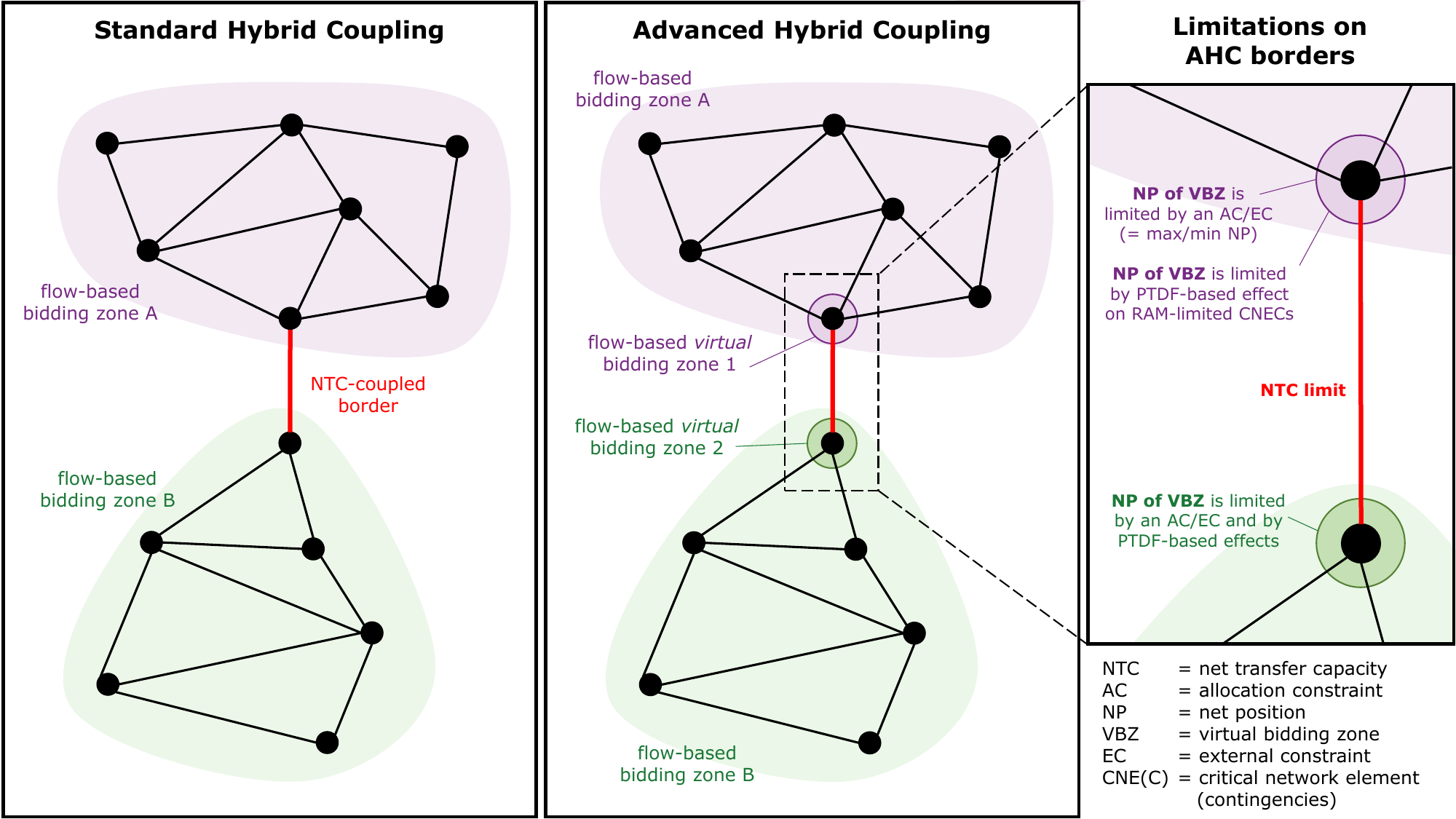}
\centering
\caption{Standard and advanced hybrid coupling}
\label{fig:explain}
\end{figure}

In the end (at the go-live of FB capacity calculation in the CWE region) a slightly modified version of “rough” hybrid coupling was adopted. In this “standard hybrid coupling” method, reservation of the CNE physical margin is not made based on the most constraining scenario of ATC transactions, but based on the forecasted (assumed) scenario (see \citealp{cwe_fbmc_approval_doc_2014}). 

At that time, it was noted that the TSOs are committed to studying and potentially implementing the advanced hybrid coupling solution. Currently, both FB regions (Core CCR and Nordic CCR) are working on implementing AHC, which should be finalized before the end of 2025 according to the latest planning\citep{entsoe_3}.

Although AHC implementation may differ in each region, the general setup of AHC and its differences to the SHC are denoted on Fig. \ref{fig:explain}:

\begin{itemize}[itemsep=0.1em, parsep=0pt, label={--}]
\item In the FB region applying SHC the impact of transactions over NTC/ATC interconnector on the CNEs in FB region is not explicitly modeled. Hence a reservation of the capacity is made based on the forecasted transactions as it will be described in the following sub-section. Transactions over the NTC/ATC interconnector are only subjected to the corresponding NTC/ATC constraint.
\item In the FB region applying AHC a virtual bidding zone (VBZ) is introduced on the connection node of the NTC/ATC interconnector. This VBZ explicitly models the impact of transactions over the NTC/ATC interconnector on the CNEs in the given FB region, using PTDF factors. Hence in AHC setup the transactions over NTC/ATC interconnector are limited by following constraints:
\begin{enumerate}
\item NTC/ATC constraint on the interconnector
\item FB constraints on the flows over individual CNEs 
\item FB allocation constraints on the VBZ net positions (this constraint is optional as it will be described in the next sub-section)
\end{enumerate}

\end{itemize}

The following sections describes the capacity calculation and capacity allocation process and the parameters that are impacted by the hybrid coupling solution.

\subsection{Impact on capacity calculation}
\label{shc_ahc:sub2}
A flow chart of the capacity calculation process is given in Fig. \ref{fig:core-process} below, which in general applies to both Core CCR and Nordic CCR.
\begin{figure}[!h]
    \centering
    \includegraphics[width=1\linewidth]{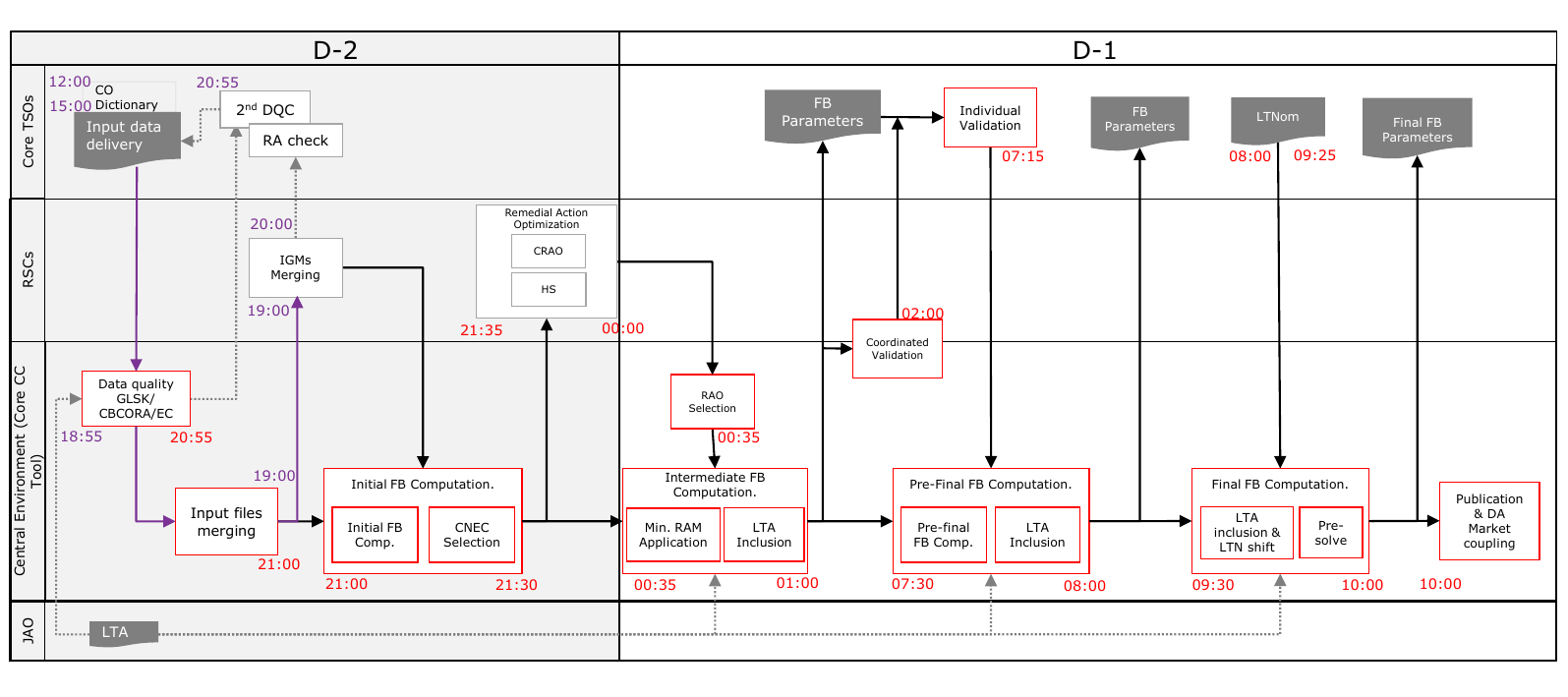}
    \caption{Capacity calculation process in Core CCR (based on \citealp{core_fbmc_workshop}, p. 47)}
    \label{fig:core-process}
\end{figure}
Following capacity calculation steps are foreseen:
\begin{enumerate}
\item Firstly, the necessary input data is delivered by TSOs to the coordinated capacity calculator, which is performing the calculation. Most important inputs from each TSO are the grid model, list of CNECs and external constraints, as well as the generation (and load) shift keys (G(L)SK) to be applied in the calculation.\footnote{Note that, since the process is started two days ahead of the delivery (D-2), the grid models contain best forecasts of the transmission grid state at the moment of delivery. Hence the grid model is also called two-days ahead congestion forecast (D2CF), and among other parameters it also includes the forecasted exchanges to neighboring CCRs.}

\item After receiving the inputs, the capacity calculation process starts with the merging of TSO individual grid models (IGM) into the common grid model (CGM). Subsequently, based on the CGM, initial flow-based parameters (PTDFs and reference flows) are calculated. In this step, CNECs that are not sensitive to cross-zonal exchanges are filtered out from the initial CNEC list delivered by TSOs. This filtering or selection is done using a PTDF-based criteria, as denoted in Eq. \ref{eq:im_ptdf_threshold}. CNECs with maximum zone-to-zone PTDF lower than a specific threshold (5\% in Core CCR and Nordic CCR) are not considered in the further calculation. This step is also impacted by the hybrid coupling concept, as due to introduction of virtual bidding zones in AHC there will be more CNECs which are considered compared to SHC.
 
\begin{equation}\label{eq:im_ptdf_threshold}
\underbrace{max(PTDF_{X,i})}_{\substack{\text{over all bidding} \\ \text{zones X in given CCR}}}- \underbrace{min(PTDF_{X,i})}_{\substack{\text{over all bidding} \\ \text{zones X in given CCR}}} \geq Threshold
\end{equation}

\item Remaining available margin ($RAM$) represents available capacity on a CNEC in a situation with no cross-zonal exchanges within the corresponding CCR. Hence it is obtained by deducting the zero-exchange flow (in the given CCR) $F_{0,CCR}$ as well as the reliability margin $FRM$ from the maximum admissible flow on each considered CNEC: 
\begin{equation}\label{eq:im_ram_init}
\overrightarrow{RAM}_{init} = \overrightarrow{F}_{max}-\overrightarrow{F}_{0,CCR}-\overrightarrow{FRM}
\end{equation}
\begin{equation}\label{eq:im_f0_ccr}
\overrightarrow{F}_{0,CCR} = \overrightarrow{F}_{ref} - \mathbf{PTDF_{CCR}}\times\overrightarrow{NP}_{ref,CCR}
\end{equation}

$F_{0,CCR}$ is obtained by zeroing out the effect of cross-zonal exchange in the given CCR, which is estimated using PTDF and net positions of the bidding zones in the given CCR. The hybrid coupling setup impacts the $F_{0,CCR}$ and consequently the initial $RAM_{init}$ as the borders to neighboring CCR can get included into this equation via virtual bidding zones (e.g. the product \(\mathbf{PTDF_{CCR}}\times\overrightarrow{NP}_{ref,CCR}\) will be different in AHC and SHC setup).

\item The last step significantly impacted by the hybrid coupling setup is the application of the minimum RAM. 
In this step, it is ensured that a minimum capacity  required by EU Regulation 2019/943 is available for cross-zonal exchanges. 
It should be noted that the minimum capacity applies to exchanges over all borders, and not just within the CCR. On the other hand, RAM represents an available margin for transactions within the CCR. Hence, to apply the minimum capacity it is necessary to take into account not only the RAM but also the impact of other exchanges which are not within the given CCR.
This is done by introducing a variable called unscheduled allocated flows $F_{uaf}$ which represents the impact of other exchanges
\begin{equation}\label{eq:im_ram70}
RAM+F_{uaf} \geq R_{amr}\cdot F_{max}
\end{equation}
and it is calculated as a difference between the in global zero-exchange flow $F_{0,all}$ and the CCR zero-exchange flow $F_{0,CCR}$, as denoted in Eq. \eqref{eq:im_fuaf}. The unscheduled allocated flow is impacted by the hybrid coupling setup through  $F_{0,CCR}$ as it was explained in the previous step (note that $F_{0,all}$ is not impacted by the hybrid coupling setup).
\begin{equation}\label{eq:im_fuaf}
F_{uaf} = F_{0,CCR}-F_{0,all}    
\end{equation}
\begin{equation}\label{eq:im_f0_all}
\overrightarrow{F}_{0,all} = \overrightarrow{F}_{ref} - \mathbf{PTDF_{all}}\times\overrightarrow{NP}_{ref,all}
\end{equation}

$R_{amr}$ in inequality \eqref{eq:im_ram70} represents the minimum RAM factor per CNEC, which generally shall be equal to 70\% starting from January 1st 2026, but currently deviates from this value based on the granted derogations and action plans (see \citealp{acer_action_plans}). 

Based on the minimum capacity requirement in \eqref{eq:im_ram70}, as well as the equations \eqref{eq:im_f0_ccr}, \eqref{eq:im_fuaf} and \eqref{eq:im_f0_all} - it is possible to derive the final $RAM$ which complies with this requirement. 

The difference between the initially calculated $RAM_{init}$ and the final RAM is called $AMR$ (Adjustment for minimum RAM), and it is either positive value (if initial RAM had to be adapted) or zero otherwise:
\begin{equation}\label{eq:im_amr}
AMR = max(RAM - RAM_{init}, 0)
\end{equation}

\begin{equation}\label{eq:im_ram_diff}
\begin{aligned} 
RAM - RAM_{init} &=  R_{amr}\cdot F_{max}-F_{uaf}-(F_{max}-F_{0,CCR}-FRM) \\ 
&=  (R_{amr}-1)\cdot F_{max}+F_{0,all}+FRM
\end{aligned}
\end{equation}

As it can be seen, the $AMR$ parameter is not impacted by the hybrid coupling setup, as it only depends from $F_{0,all}$ which is not impacted.

However, note that the Core CCR applies an additional requirement that the minimum available margin shall be higher than 20\% for CCR-internal transactions. 
\begin{equation}\label{eq:im_core20}
RAM \geq 20\% \cdot F_{max}
\end{equation}

This has an impact on the equation \eqref{eq:im_ram_diff} that now becomes:
\begin{equation}\label{eq:im_ram_init_core20}
RAM - RAM_{init} = max
\begin{pmatrix}
     (R_{amr}-1)\cdot F_{max}+F_{0,all}+FRM,\\
    20\%\cdot F_{max}-(F_{max}-F_{0,CCR}-FRM)
\end{pmatrix}
\end{equation}

When considering the CCR-specific requirement for minimum RAM, the hybrid coupling setup could have some impact on the AMR, as it becomes dependent on $F_{0,CCR}$. Generally, requirement \eqref{eq:im_ram70} is stricter than requirement \eqref{eq:im_core20}.\footnote{If we assume that \(R_{amr}=70\%\) this implies that requirement \eqref{eq:im_core20} is only more strict when \(F_{uaf} \geq 50\%\cdot F_{max}\). Due to the way CCRs are delimited, it is expected that this will rarely occur on any CNEC. It can only occur, if there are borders left using SHC.}
\end{enumerate}

Capacity calculation contains further steps such as LTA inclusion (long-term allocations), non-costly remedial actions optimization (NRAO), and TSO domain validation. Although these steps impact the final FB parameters (PTDF and RAM), they have a rather weak interdependency with the hybrid coupling setup, at least on the conceptual level. Consequently, they are not further elaborated in this section.

Furthermore, the use of External Constraints (EC) could be impacted by the Hybrid Coupling approach. In AHC setup, virtual bidding zones external constraints (VBZ ECs) can be used to denote the maximum and minimum net position of the virtual bidding zone in the given CCR, which corresponds to maximum import/export over the given border to neighboring CCR. 

Note that although VBZ EC is a redundant parameter to NTC limit (ideally these should have the same value), as depicted in Fig. \ref{fig:explain}, it makes a difference from a process perspective, because the NTC value is not known to the FB capacity calculation process of the given CCR. Hence the main benefit of using VBZ ECs is an increased accuracy of FB domain indicators such as minimmum/maximum net positions of bidding zones, as well as the min/max bilateral exchanges. 

\subsection{Impact on capacity allocation}
\label{shc_ahc:sub3}

\begin{figure}[!h]
    \centering
    \includegraphics[width=0.7\textwidth]{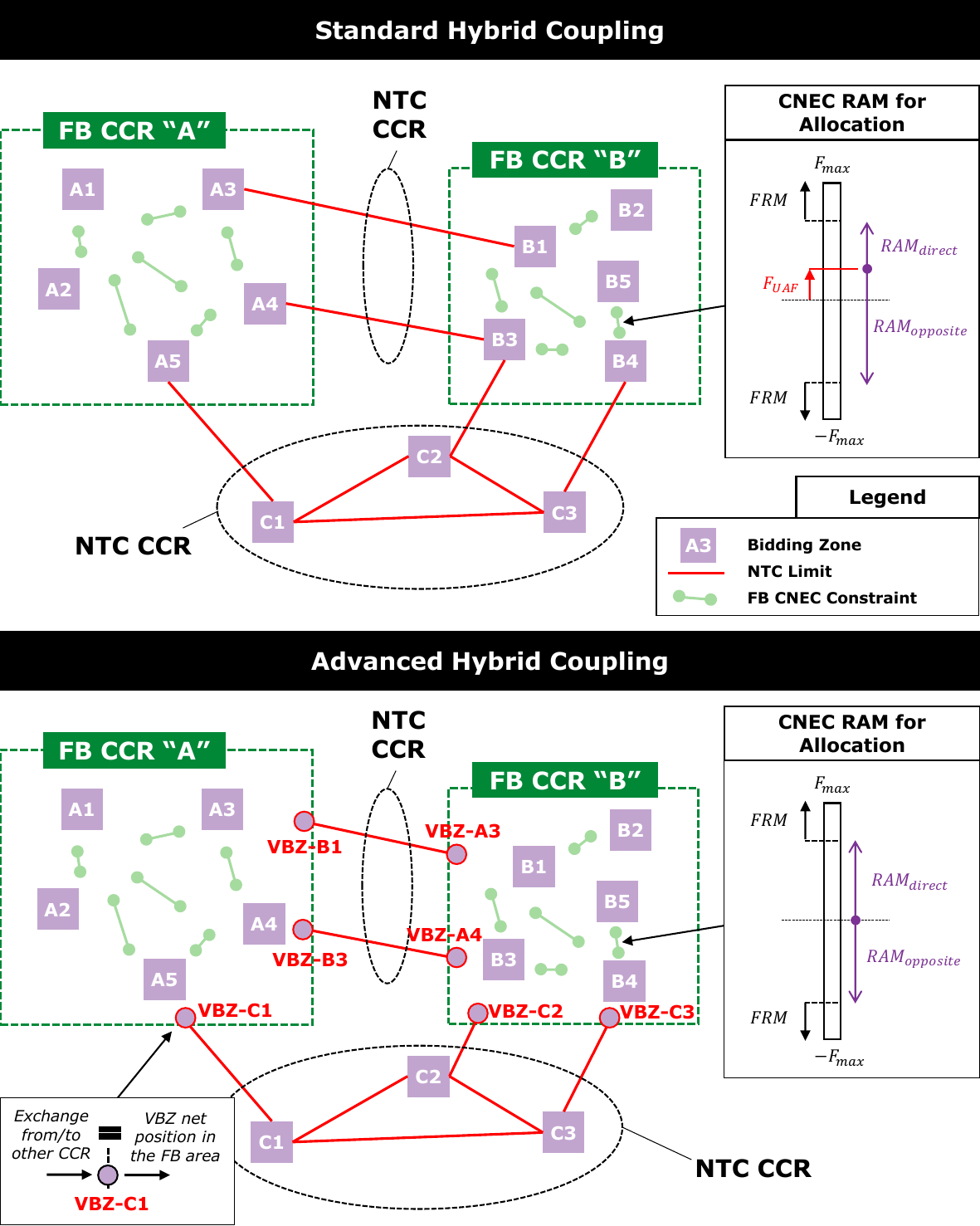}
    \caption{Capacity allocation and CNEC RAM in standard and advanced hybrid coupling.}
    \label{fig:detailed_AHC}
\end{figure}

In the European day-ahead market, cross-zonal capacity is allocated in the market coupling process. During market coupling, “buy” and “sell” orders for electricity from different bidding zones are matched in an optimized way, considering the available cross-zonal capacity. 
Cross-zonal capacity constraints are given either as an NTC limit per bidding zone border, or as a set of flow-based constraints for the respective CCR. This is depicted in Fig. \ref{fig:detailed_AHC}, in which two regions are applying the flow-based approach and the other two are applying NTC limitations\footnote{Note that one NTC region just consists of interconnections between two FB regions "A" and "B", while the other contains borders between bidding zones C1-C3 and interconnections to the mentioned FB regions.}.

The NTC constraint applies to and limits the exchanges $EX_{X\rightarrow Y}$ over the respective interconnector and direction X\textrightarrow Y
\begin{equation}\label{eq:im_ntc_limit}
EX_{X\rightarrow Y} \leq NTC_{X\rightarrow Y}
\end{equation}

Flow-based constraints are given as a set of linear equations, and they are limiting exchanges on all borders between bidding zones in the given region (for example: in CCR "A")
\begin{equation}\label{eq:im_ptdf_limit}
\begin{bmatrix}
PTDF^{A_1\rightarrow A_2}_{CNEC 1} & ... & PTDF^{A_i\rightarrow A_j}_{CNEC 1}\\
... & ... & ...\\
PTDF^{A_1\rightarrow A_2}_{CNEC X} & ... & PTDF^{A_1\rightarrow A_2}_{CNEC X}
\end{bmatrix} \times 
\begin{bmatrix}
EX^{A_1\rightarrow A_2} \\ ...\\EX^{A_i\rightarrow A_j}
\end{bmatrix} \leq
\begin{bmatrix}
RAM_{CNEC 1} \\ ...\\RAM_{CNEC X}
\end{bmatrix}
\end{equation}

As it can be seen in Inequality \eqref{eq:im_ptdf_limit}, only the impact of exchanges within the flow-based region ($EX^{A_1\rightarrow A_2}$,… ,$EX^{A_i \rightarrow A_j}$) on CNECs is explicitly considered during the allocation phase. 
Hence, in the standard hybrid coupling setup the impact of the exchanges to other CCRs is implicitly modeled by adjusting the RAM value to account for the forecasted exchange, as depicted in Figure \ref{fig:detailed_AHC} and described in Section \ref{shc_ahc:sub2}. The following should be noted:
\begin{itemize}[itemsep=0.1em, parsep=0pt, label={--}]
    \item CNEC RAM is reduced in one direction while it is increased in the opposite, according to the forecasted exchanges. 
    \item To ensure that a potential forecast error will not lead to exceeding the maximum capacity, FRM value must be dimensioned accordingly.
\end{itemize} 

In the advanced hybrid coupling setup, virtual bidding zones are introduced on the ends of NTC interconnectors to other CCRs, as denoted in Fig. \ref{fig:detailed_AHC}. They are used to explicitly model the impact of exchanges to other CCRs on the flow-based parameters. In order to enable this, virtual bidding zones have following characteristics:
\begin{itemize}[itemsep=0.1em, parsep=0pt, label={--}]
    \item Virtual bidding zones do not have any generation or load. Their total net position is always zero. Note that the \emph{total} net position of a VBZ is a sum of two component: One is the exchange from/to other CCR, over the NTC interconnector(s), and the other is exchange towards the flow-based region. In this way, exchange from/to other CCR is propagated into FB constraints (as these two components are equal).  
\begin{equation}\label{eq:im_np_vbz1}
NP_{VBZ,total} = EX_{other CCR\rightarrow VBZ} - EX_{VBZ\rightarrow FB area} = 0
\end{equation}
\begin{equation}\label{eq:im_np_vbz2}
EX_{other CCR\rightarrow VBZ} = EX_{VBZ\rightarrow FB area}
\end{equation}
\item Virtual bidding zones are represented in the PTDF matrix in the same way as normal bidding zones. As  virtual bidding zones are located on the ending node(s) of the corresponding interconnector to another CCR, GSKs are assigned to those nodes based on the expected (forecasted) flow distribution. 
\begin{equation}\label{eq:im_ptdf_limit_2}
\begin{bmatrix}
PTDF^{A_1\rightarrow A_2}_{CNEC 1} & ... & PTDF^{A_i\rightarrow A_j}_{CNEC 1}& \mathcolor{red}{PTDF^{VBZ_n\rightarrow A_m}_{CNEC 1}}\\
... & ... & ... & \mathcolor{red}{...}\\
PTDF^{A_1\rightarrow A_2}_{CNEC X} & ... & PTDF^{A_1\rightarrow A_2}_{CNEC X}& \mathcolor{red}{PTDF^{VBZ_n\rightarrow A_m}_{CNEC X}}
\end{bmatrix} \times 
\begin{bmatrix}
EX^{A_1\rightarrow A_2} \\ ...\\EX^{A_i\rightarrow A_j}\\ \mathcolor{red}{EX^{VBZ_n\rightarrow A_m}}
\end{bmatrix} \leq
\begin{bmatrix}
RAM_{CNEC 1} \\ ...\\RAM_{CNEC X}
\end{bmatrix}
\end{equation}

\end{itemize} 

Since the impact of exchanges to other CCRs on CNECs is explicitly considered in the AHC setup, it is not necessary to have any adjustment of the RAM based on the forecasted flow. These exchanges now compete for the scarce CNEC capacity (e.g. RAM) with other exchanges in the flow-based region. 
Therefore, in the AHC setup the full CNEC capacity is allocated based on the market algorithm, i.e. maximizing social welfare while ensuring operational security.
With regards to capacity allocation, advanced hybrid coupling offers several advantages leading to higher efficiency than standard hybrid coupling solution:
\begin{enumerate}
\item \textbf{AHC eliminates "inefficiencies" through forecast errors}: There is no need for "ex-ante" reservation of CNEC capacity, which is prone to forecasting errors. 
\item \textbf{AHC ensures optimal and fair allocation of CNEC capacities between different CCRs}: In the AHC setup all exchanges (incl. those to other CCRs) compete for scarce CNEC capacity, which ensures non-discrimination (hence increasing fairness) and maximizes social welfare.
\end{enumerate} 

\subsection{Summary}
\label{shc_ahc:sub4}
As discussed in Section \ref{shc_ahc:sub2}, the choice of the hybrid coupling setup will have an impact on many parameters in the capacity calculation. The most important are summarized here:
\begin{itemize}[itemsep=0.1em, parsep=0pt, label={--}]
\item \textbf{CNEC selection}: The introduction of virtual bidding zones in AHC setup could lead to a higher number of CNECs in the capacity calculation, as these are determined based on the maximum zone-to-zone PTDF.
\item \textbf{Change in FB capacities through RAM values}: In the SHC setup CNEC capacity is reduced in one direction and increased in the other direction based on the forecasted exchanges to other CCRs. In AHC there is no capacity reservation made, e.g. the RAM is determined for the state of zero-exchange to other CCRs.
\item \textbf{Minor impact on the minimum RAM requirement (if any)}: The hybrid coupling setup is not impacting the 70\%-minRAM requirement. There is some effect on the CCR-specific minRAM (such as 20\% in Core). However, the 70\%-rule is the generally stricter requirement.
\item \textbf{Allocation constraints}: They can be used in the AHC setup to denote maximum import/export on the respective border to neighboring CCR. This enhances the accuracy of the FB domain and related indicators (such as min/max net positions and bilateral exchanges, as well as presolved CNECs) 
\end{itemize}

Furthermore, the hybrid coupling setup has significant implications for capacity allocation, as described in Section \ref{shc_ahc:sub3}, with following main differences:
\begin{itemize}[itemsep=0.1em, parsep=0pt, label={--}]
\item \textbf{Efficiency}: AHC eliminates the need for "ex-ante" reservation of CNEC capacities, which is prone to forecast errors and therefore results in inefficiencies. These are removed through AHC.
\item \textbf{Fairness}: In the AHC setup all exchanges compete for scarce CNEC capacity based on the market algorithm, while in SHC CNEC capacity is reserved based on forecasted exchanges to other CCRs. 
SHC therefore prioritizes exchanges to other CCRs compared to the given FB CCR, while AHC ensures fair competition leading to maximal social welfare.
\end{itemize}

\newpage
\part{Model-based evaluation how advanced hybrid coupling affects market results and congestion management}
\section{Setup of the study}
\label{sec:setup}

The study is based on the setup of \cite{schonheit2021toward}, which fully describes the model, data and assumptions. The grid for the study at hand remains the same. This section describes how the study setup deviates from \cite{schonheit2021toward} to model the effects of advanced hybrid coupling, in terms of two-days ahead uncertainties and generating capacities (\ref{setup:sub1}), modeling of flow-based market coupling parameters (\ref{setup:sub2} and \ref{setup:sub3}), modeling of trade restrictions (\ref{setup:sub4}) and model assumptions (\ref{setup:sub5}). The full model is detailed in Appendix \ref{app:model}. It is largely similar to the model in \cite{schonheit2021toward}. This section's purpose is to highlight the most important deviations. 

\subsection{Two-days ahead uncertainty and generating units}
\label{setup:sub1}

As outlined in Section \ref{sec:shc_ahc} (muss ergaenzt werden), flow-based market coupling  is subjected to D-2 uncertainties, SHC more so than AHC. The reason for this being that D2CF contains forecasted amounts of exchange across the NTC-coupled borders. These lead to a utilization of CNEs due to exchanges, the so-called $f_{t,l}^{\mathrm{uaf}}$ (unscheduled allocated flows, see \citealp{entsoe_1}). The mathematical details follow below. Along with uncertain information on load as well as generation levels of conventional and renewable energy sources, these predicted exchanges and their resulting $f_{t,l}^{\mathrm{uaf}}$ constitute the main uncertainties in the two days-ahead forecasts.

\begin{figure}[h!]
\includegraphics[width=0.95\textwidth]{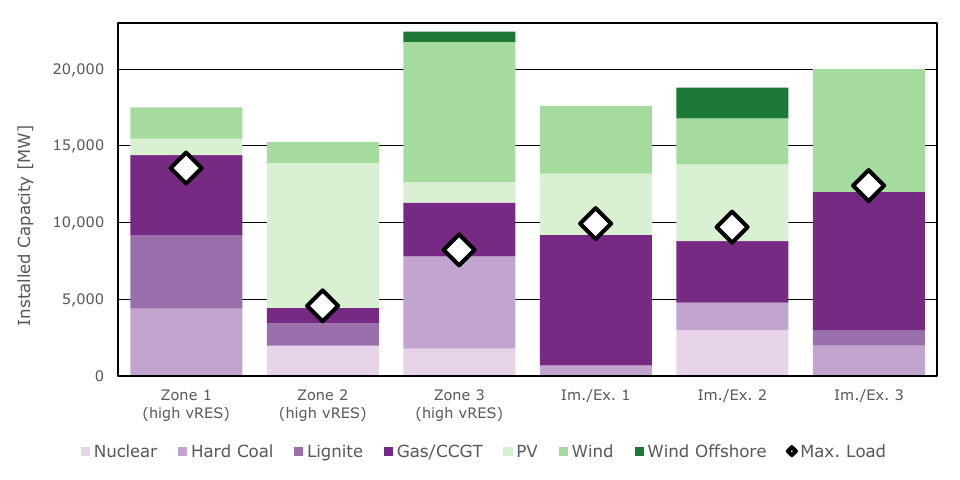}
\centering
\caption{Installed generating capacity by zone and fuel-type. Zone 1-3 are taken from the scenario "high variable renewable energy sources" ("high vRES") from \cite{schonheit2021toward}.}
\label{fig:generation}
\end{figure}

Hence, to introduce uncertainty into the base case (D2CF) in this study, the generation structure of the adjacent, non-FB zones is changed, as depicted in Fig. \ref{fig:generation}. The generating units of the three FB zones is identical to the scenario "high variable renewable energy sources" ("high vRES")  of \cite{schonheit2021toward}. Further, the time series for the availability factors of renewable energy sources are subjected to uncertainty. This is done by multiplying the original time series with a random factor, normally distributed around a mean of one and a standard deviation of 0.2 for RES located in FB zones and 0.3 for RES located in non-FB zones. The latter, higher standard deviation resembles the higher level of uncertainty contained in the forecasted exchange across the NTC-coupled borders.

The base case (\ref{subsec:mod_d-2}) uses $ren_{t,n}^\mathrm{D-2}$ as a renewable forecast. The uncertainty is removed in the D-1 market coupling (\ref{subsec:mod_d-1shc} and \ref{subsec:mod_d-1ahc}) and congestion management model (\ref{subsec:mod_d-0}), which use the original time series $ren_{t,n}$ without fuzziness.

\subsection{Changes in flow-based parameters due to advanced hybrid coupling}
\label{setup:sub2}

\begin{figure}[h!]
\includegraphics[width=0.95\textwidth]{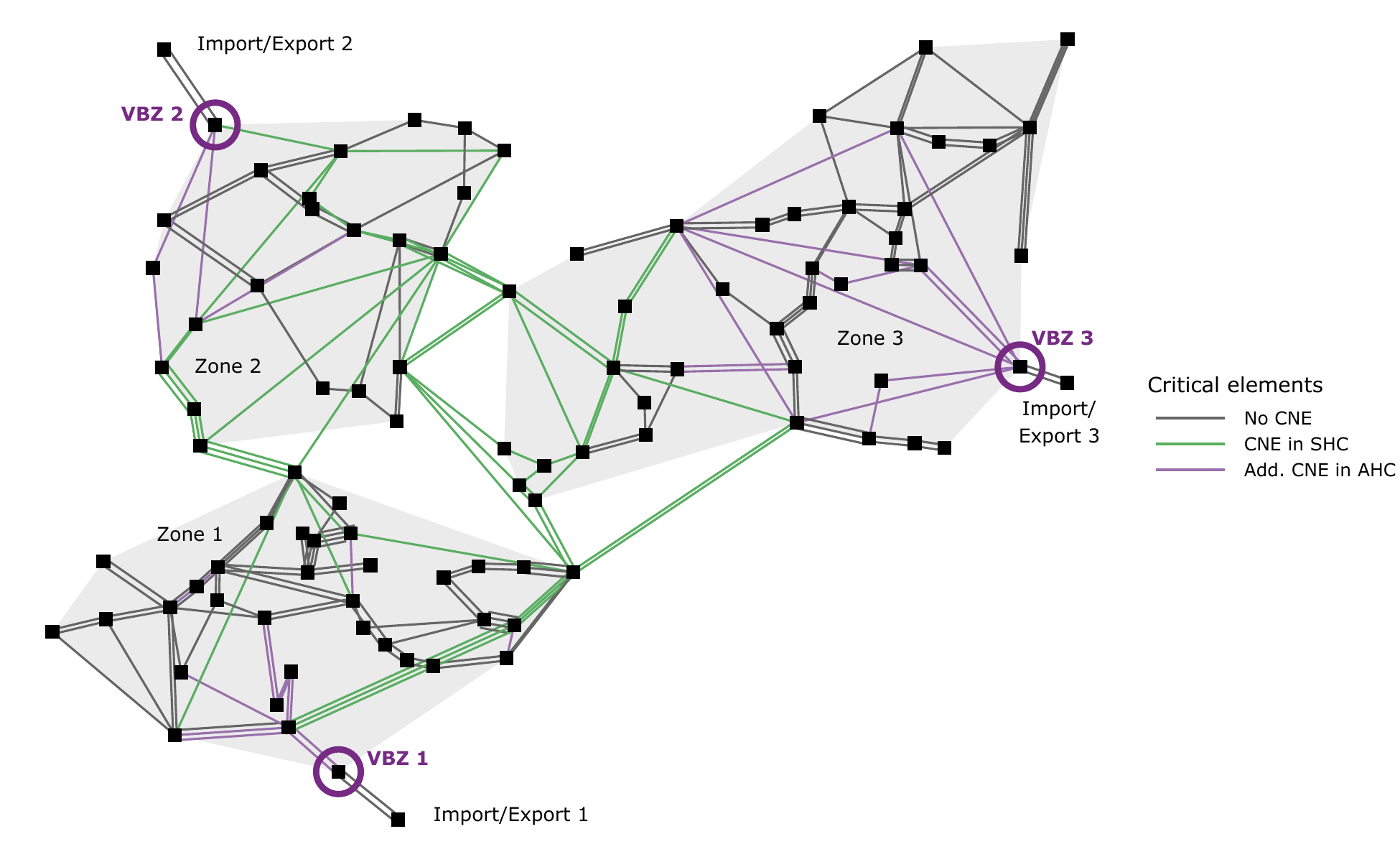}
\centering
\caption{Power grid of the study. The bidding zones are three flow-based zones 1-3 and three non-flow-based zones ("Import/Export" 1-3). With advanced hybrid coupling (AHC), three virtual bidding zones (VBZs) are introduced. The colors of the lines depict the selected critical network elements in standard hybrid coupling (SHC) and AHC.}
\label{fig:grid}
\end{figure}

Another important change, necessary to model the differences between SHC and AHC, is the introduction of virtual bidding zones (VBZs). Fig. \ref{fig:grid} highlights how the end nodes (within the FB zones) of the NTC-coupled interconnectors represent the VBZs 1-3. 

The introduction of VBZs is represented mathematically by extending the GSK matrix: Concretely, in this study three columns are added to $\mathbf{G}$, representing the three additional zones, resulting in $\mathbf{G^\mathrm{AHC}}$. The three nodes that represent the three VBZ, each get a GSK factor of 1 in the respective column.\footnote{Due to the simple topology in this study, each VBZ only consists of one node, with a GSK factor of 1. In reality, a more complex topology can lead to a multi-node VBZ. These can receive different GSK factors, representing an assumption on the distribution of the imports/exports on the NTC-coupled border across nodes. Mathematically, it only makes a difference, if these two nodes have different node-to-line PTDF values, meaning a different effect on the meshed grid.} $\mathbf{G^\mathrm{AHC}}$ can then be multiplied with the nodal PTDF matrix, resulting in a zonal PTDF matrix with three additional columns/zones. 

Having more zones also means comparing more zone-to-zone combinations when selecting CNEs. Moving from SHC to AHC in this test model, not only $\frac{3 \cdot 2}{2} = 3$ but $\frac{6 \cdot 5}{2} = 15$ combinations are tested due to the increase in FB zones from three to six. More CNEs will fulfil the maximum zone-to-zone PTDF criterion of 5\% \citep{entsoe_1}, so in this study it leads to an increase in CNEs from 62 (SHC) to 95 (AHC). The result is also visualized in Fig. \ref{fig:grid}. Naturally, the lines closest to the VBZs are \emph{additionally} selected when moving from SHC to AHC. This allows for a more accurate quantification of the VBZs' effect on the grid, introduced as constraints for cross-border trade during market coupling.

Also, deviating from \cite{schonheit2021toward}, this study computes "n-1"-secure market domains as well as "n-1"-secure redispatch. This is done by computing LODF (line outage distribution factors), see \cite{guo2009direct}. These indicate what fraction of the flow on line $l_1$ is transported on $l_2$ when $l_1$ becomes unavailable. LODFs alter the nodal and zonal PTDF values. During market coupling, the five worst outages are considered (the five highest absolute LODF-values for the respective line) for each CNE, turning them in to CNECs, in addition to the "n-0" case. Thus, the amount of CNECs  $|J| = 6 \ \cdot$ (amount of CNEs). During congestion management, the two worst outages are considered in addition to the "n-0" case. Thus, $|L^\mathrm{c}| = 3 \cdot |L|$

\subsection{Unscheduled allocated flows (UAF) and remaining available margin (RAM) in standard and advanced hybrid coupling}
\label{setup:sub3}

To complement the description of the setup, it is worth mathematically outlining the main difference between SHC and AHC in terms of remaining available margins, following \cite{entsoe_1} and \cite{entsoe_3}.

When the D2CF is computed based on D-2 forecasts, there is a resulting reference flow on each element/line, $f_{t,l}^{\mathrm{D-2,ref}}$. To compute the flows in absence of trade activities \emph{between flow-based zones}, the D-2 net positions of the FB zones, contained in $\mathbf{n}_t^\mathrm{FB,D-2}$, are multiplied by the zonal PTDFs and the resulting flows are subtracted from the reference flows, as shown in Eq. \ref{eq:f0fb}. However, the resulting $f_{t,l}^{\mathrm{0,FB}}$ still contains the flows caused by exchanges across NTC-coupled borders.

\begin{equation}
    \mathbf{f}_t^{\mathrm{0,FB}} = \mathbf{f}_t^{\mathrm{D-2,ref}} - \mathbf{P^\mathrm{\mathrm{Z}}} \cdot \mathbf{n}_t^\mathrm{FB,D-2} \label{eq:f0fb}
\end{equation}

To get the so-called $f_{t,l}^{\mathrm{0,all}}$, a counterfactual state of zero trade across \emph{all} zones called "zero balance", the D-2 \emph{global} net positions of the flow-based \emph{and} the non-flow-based zones are multiplied with the zonal PTDFs and subtracted from the reference flow, see Eq. \ref{eq:f0all}. The equation splits up the flow-based and non-FB terms for clarity. The exchanges ($EX_{t,z}^\mathrm{D-1}$: exports from non-FB into a FB zone across an NTC-coupled border) are contained in $\mathbf{x}_{t}^\mathrm{D-2}$. Neglecting losses, these can be multiplied by the nodal PTDF values of the NTC-connected nodes, $\mathbf{P^\mathrm{N^{nFB}}}$.

\begin{equation}
    \mathbf{f}^{\mathrm{0,all}} = \mathbf{f}^{\mathrm{D-2,ref}} - \mathbf{P^\mathrm{\mathrm{Z}}} \cdot \mathbf{n}_{t}^\mathrm{global,D-2} -  \mathbf{P^\mathrm{N^{nFB}}} \cdot \mathbf{x}_{t}^\mathrm{D-2} \label{eq:f0all}
\end{equation}

Importantly, the difference between the flows without trade activity between FB zones and the flows without \emph{any} trade activity is the above-mentioned $f_{t,l}^{\mathrm{uaf}}$ (Eq. \ref{eq:fuaf}). This values directly depends on the assumption made regarding the amount of trade across the NTC-coupled border.

\begin{equation}
     f_{t,l}^{\mathrm{uaf}} = f_{t,l}^{\mathrm{0,FB}} - f_{t,l}^{\mathrm{0,all}} \label{eq:fuaf} 
\end{equation}

\begin{figure}[h!]
\includegraphics[width=0.95\textwidth]{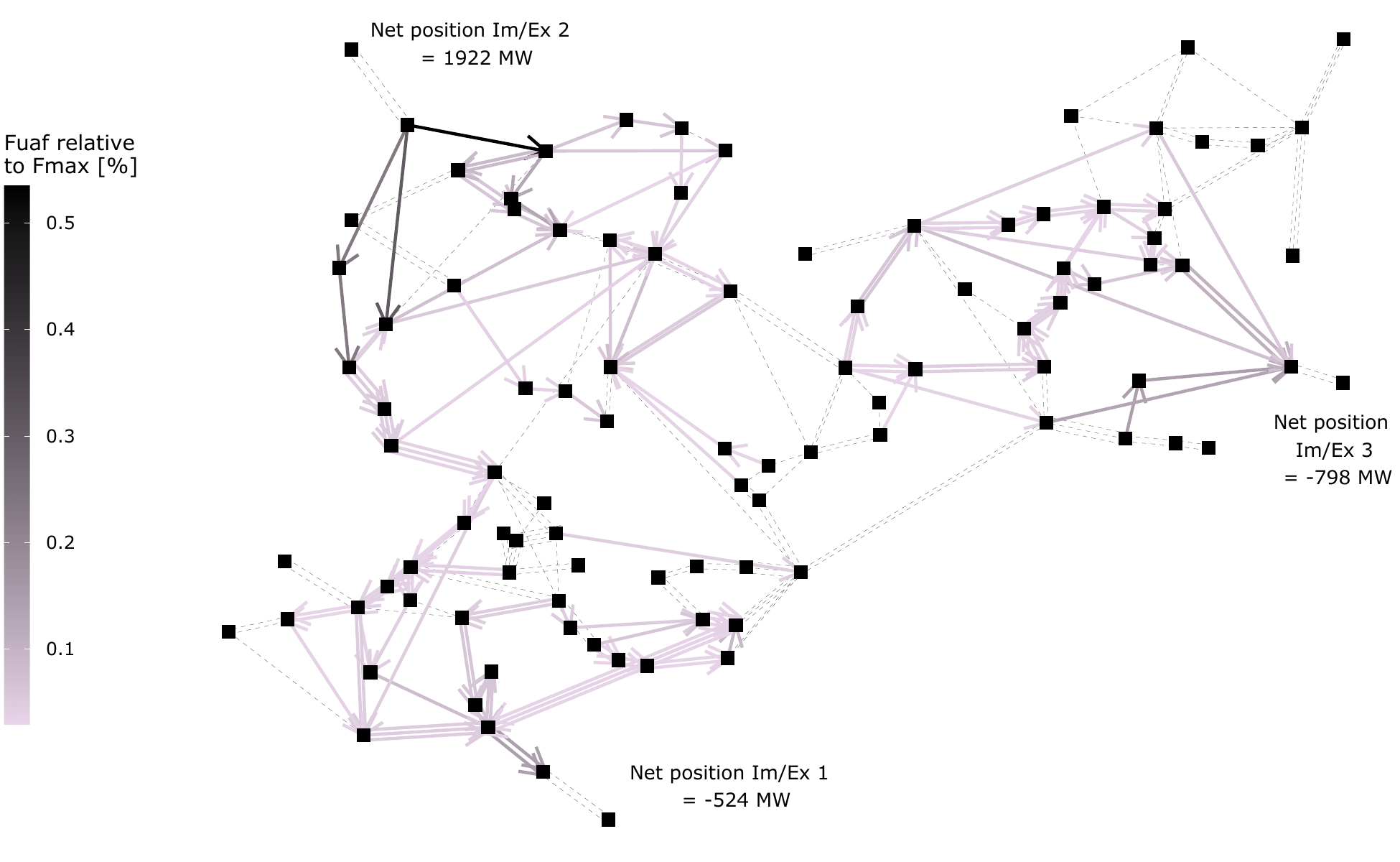}
\centering
\caption{Example of $f_{t,l}^{\mathrm{uaf}}$ with flow direction resulting from D-2 forecasts (timestamp = 9). Only lines that are within the flow-based area and have a  $|f_{t,l}^{\mathrm{uaf}}| \geq 3\%$ are included in the color scale and displayed as arrows. The remaining lines are shown as black dashed connections.}
\label{fig:fuaf_example}
\end{figure}

To complement the mathematical description, Eq. \ref{eq:ramposshc} to \ref{eq:ramnegahc} describe how the remaining available margins (RAMs), provided to flow-based market coupling as line-dependent cross-zonal trade capacities, are computed. In this study, positive RAMs are positive and negative RAMs are negative.\footnote{Other systems use positive RAMs exclusively and model the effect of negative flows by using the inverting the sign of PTDF values.} 

These are margins made available for trade between flow-based bidding zones. Crucial to note is that with SHC only $f_{t,l}^{\mathrm{0,FB}}$ is deducted from the maximum line capacity (along with a flow reliability margin). Thus, $f_{t,l}^{\mathrm{uaf}}$ is "reserved" with SHC and not made available to market activities between flow-based zones. With AHC, however,  $f_{t,l}^{\mathrm{0,all}}$ is deducted, allowing the RAM to encompass the state of no trade.

\begin{equation}
     ram_{l,t}^{\mathrm{pos, SHC}} = fmax_{l} - frm_{l} - f_{t,l}^{\mathrm{0,FB}} \label{eq:ramposshc} 
\end{equation}

\begin{equation}
     ram_{l,t}^{\mathrm{neg, SHC}} = -fmax_{l} + frm_{l} - f_{t,l}^{\mathrm{0,FB}} \label{eq:ramnegshc} 
\end{equation}

\begin{equation}
     ram_{l,t}^{\mathrm{pos, AHC}} = fmax_{l} - frm_{l} - f_{t,l}^{\mathrm{0,all}} \label{eq:ramposahc} 
\end{equation}

\begin{equation}
     ram_{l,t}^{\mathrm{neg, AHC}} = -fmax_{l} + frm_{l} - f_{t,l}^{\mathrm{0,all}} \label{eq:ramnegahc} 
\end{equation}

Finally, the criterion of minimum trading capacities ("minRAMs") has to be considered. For this study, the full 70\% of the Clean Energy Package are assumed (specifically the EU Regulation 2019/943, see \citealp{union2019regulation}). The calculation of the adjustment for minRAM (AMR) and the re-computation of RAMs are shown in Eq. \ref{eq:amrpos} - \ref{eq:ramnegahc_}. Eq. \ref{eq:amrpos} is re-formulated twice, from the formulation in the Core Day-Ahead Capacity Calculation Methodology \citep{entsoe_1} to a more intuitive version. The final line in Eq. \ref{eq:amrpos} ($0.7 \cdot fmax_{l} + frm_{l} + f_{t,l}^{\mathrm{0,all}} - fmax_{l}$) simply means that if the "demands" on the line's capacity, comprised of a) the minRAM criterion, b) the FRM and c) the already allocated flow at zero balance \emph{surpasses} the line's total capacity, the difference is added as an AMR to the RAM. The reverse logic is applied to the negative RAM (Eq. \ref{eq:amrneg})

\begin{equation}
\begin{split}
     amr_{l,t}^{\mathrm{pos}} & = max\Big[0.7 \cdot fmax_{l} - f_{t,l}^{\mathrm{uaf}} - (fmax_{l} - frm_{l} -  f_{t,l}^{\mathrm{0,FB}}), 0\Big] \\
     & = max\Big[0.7 \cdot fmax_{l} - f_{t,l}^{\mathrm{uaf}} - ram_{l,t}^{\mathrm{pos, SHC}}, 0\Big] \\
     & = max\Big[0.7 \cdot fmax_{l} + frm_{l} + f_{t,l}^{\mathrm{0,all}} - fmax_{l}, 0\Big] \label{eq:amrpos} 
\end{split}
\end{equation}

\begin{equation}
     amr_{l,t}^{\mathrm{neg}} = min\Big[-0.7 \cdot fmax_{l} - frm_{l} + f_{t,l}^{\mathrm{0,all}} + fmax_{l}, 0\Big] \label{eq:amrneg} 
\end{equation}

\begin{equation}
    \overline{ram}_{l,t}^{\mathrm{pos, SHC}} = ram_{l,t}^{\mathrm{pos, SHC}} + amr_{l,t}^{\mathrm{pos}} \label{eq:ramposshc_} 
\end{equation}

\begin{equation}
    \overline{ram}_{l,t}^{\mathrm{neg, SHC}} = ram_{l,t}^{\mathrm{neg, SHC}} + amr_{l,t}^{\mathrm{neg}} \label{eq:ramnegshc_} 
\end{equation}

\begin{equation}
    \overline{ram}_{l,t}^{\mathrm{pos, AHC}} = ram_{l,t}^{\mathrm{pos, AHC}} + amr_{l,t}^{\mathrm{pos}} \label{eq:ramposahc_} 
\end{equation}

\begin{equation}
    \overline{ram}_{l,t}^{\mathrm{neg, AHC}} = ram_{l,t}^{\mathrm{neg, AHC}} + amr_{l,t}^{\mathrm{neg}} \label{eq:ramnegahc_} 
\end{equation}

This analysis applies a minimum value for $\overline{ram}_{l,t}^{\mathrm{pos, SHC}}$ and $\overline{ram}_{l,t}^{\mathrm{neg, SHC}}$ of 20\% (see Eq. \ref{eq:im_ram_init_core20}). This ensures a minimum RAM-level for exchanges in the FB CCR.

\begin{figure}[h!]
\includegraphics[width=0.95\textwidth]{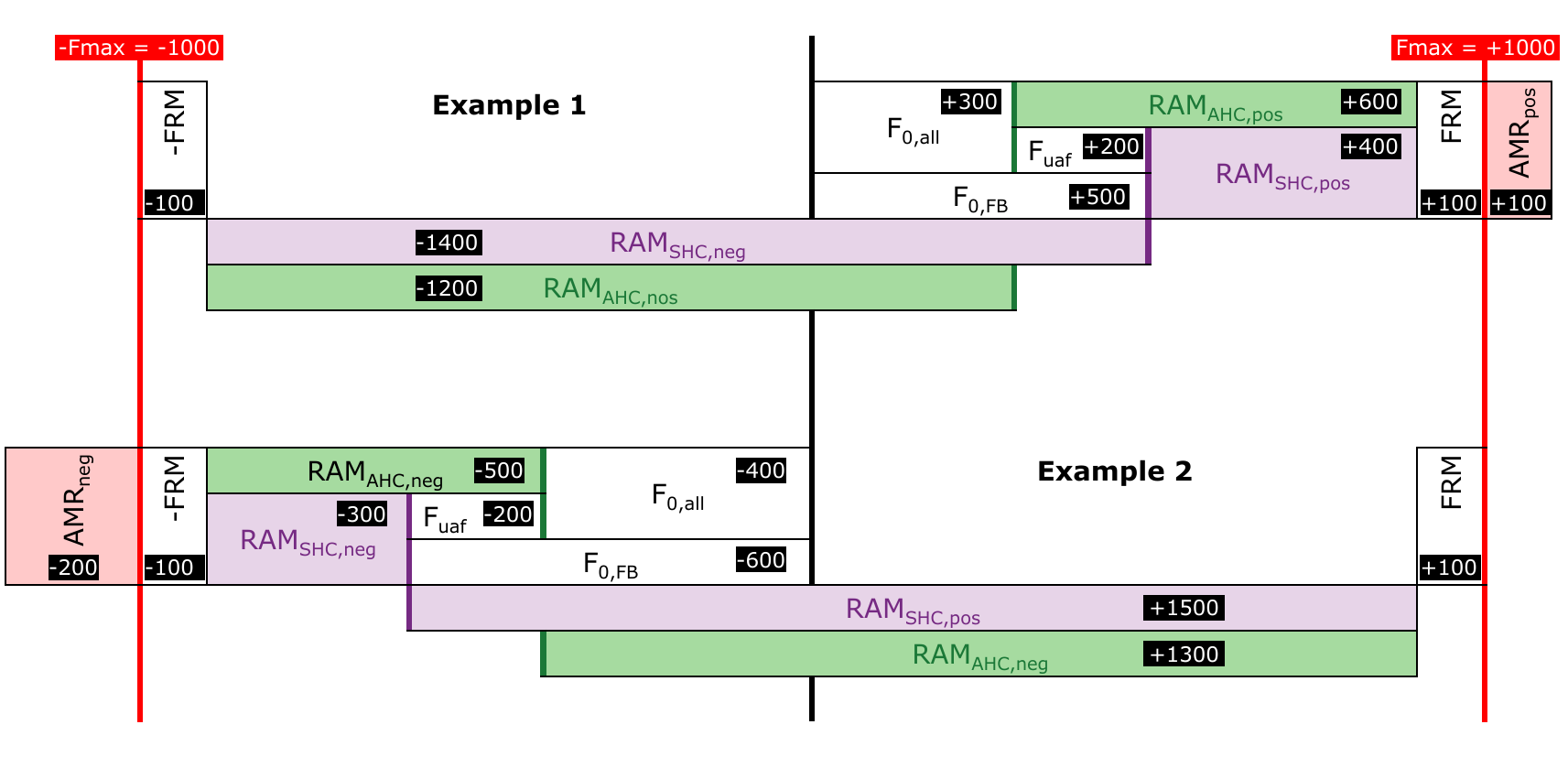}
\centering
\caption{Exemplary computation of RAMs and AMRs.}
\label{fig:amr}
\end{figure}

Summing everything up, Fig. \ref{fig:amr} gives two examples of how the AMR is computed. The RAMs for SHC and AHC are depicted as well. It becomes evident that the $f_{t,l}^{\mathrm{uaf}}$ "blocks" a certain part of the line's capacity. From a welfare-maximizing standpoint this is can lead to inefficiencies for two reasons:

\begin{enumerate}
    \item \textbf{UAFs are forecasts}. $f_{t,l}^{\mathrm{uaf}}$ is based on D-2 information and assumptions that can be erroneous, both in size and sign.  
    \item \textbf{The concept of UAF and SHC gives priority to trade on NTC-coupled borders}. $f_{t,l}^{\mathrm{uaf}}$ alters the RAM available for trade between FB zones. If, e.g., it has the same sign as $f_{t,l}^{\mathrm{0,all}}$ it can severely limit the possibility for FB trades that result in flows in the same direction of burden. However, FB trades could lead to an overall higher welfare if they were given the fraction of the line's capacity that is reserved for $f_{t,l}^{\mathrm{uaf}}$.
\end{enumerate}

This changes with AHC. Whatever forecasts are provided for exchanges on the NTC-coupled borders, their effects are reduced by computing RAMs based on $f_{t,l}^{\mathrm{0,all}}$. The "AHC RAM" (Eq. \ref{eq:ramposahc_} and \ref{eq:ramnegahc_}) renders possible for the mathematical optimization of the market coupling algorithm to determine, how a line's RAM is used for trade, allowing for the possibility of augmenting trades across FB zones and decreasing trade across NTC-coupled borders, if it leads to welfare gains.

\subsection{Restrictions for cross-zonal trade in standard and advanced hybrid coupling}
\label{setup:sub4}

The study consists of computing the D2CF according to Appendix \ref{subsec:mod_d-2}. Then, market coupling results are retrieved for the SHC and AHC setting, based on the model in Appendix \ref{subsec:mod_d-1shc} and \ref{subsec:mod_d-1ahc}. Finally, the overloads after market coupling are resolved through congestions management, see Appendix \ref{subsec:mod_d-0}. Fig. \ref{fig:process_d_2_1_0} provides a summary of the computation steps necessary for the analysis.

\begin{figure}[h!]
\includegraphics[width=0.7\textwidth]{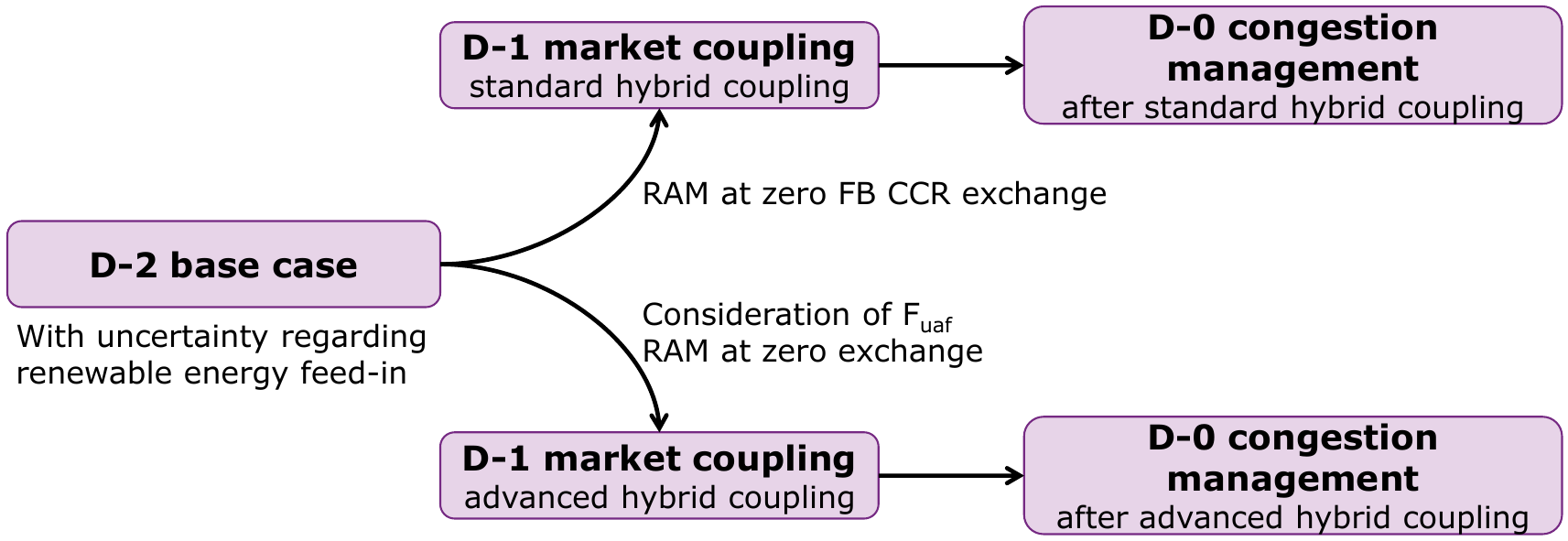}
\centering
\caption{There is a common D-2 base case for both runs, SHC and AHC, which is subjected to uncertain forecasts for renewable feed-in. Different RAMs are considered during market coupling as detailed in Section \ref{setup:sub3}. Each market coupling is followed by a congestion management analysis.}
\label{fig:process_d_2_1_0}
\end{figure}

The way that cross-zonal trade is limited differs between standard and advanced hybrid coupling:

\begin{itemize}[itemsep=0.1em, parsep=0pt, label={--}]
    \item \underline{Standard hybrid coupling}:
    \begin{itemize}[itemsep=0.1em, parsep=0pt, label={--}]
        \item \textbf{NTCs}: The export/import from/to non-flow-based zones is limited by NTC-values, see Eq. \ref{eq:modfb:ex2} and \ref{eq:modfb:ex3}
        \item \textbf{PTDF-based net position effect on CNECs}: The net positions of the \emph{physical} flow-based zones (zone 1-3) are limited by their PTDF-based effects on the CNECs' RAMs, see Eq. \ref{eq:modfb:rampos} and \ref{eq:modfb:ramneg}
    \end{itemize}
    \item \underline{Advanced hybrid coupling}:
    \begin{itemize}[itemsep=0.1em, parsep=0pt, label={--}]  
        \item \textbf{NTCs}: The export/import from/to non-flow-based zones is limited by NTC-values, see Eq. \ref{eq:modfb:ex2ahc} and \ref{eq:modfb:ex3ahc}. Because the net position of the virtual bidding zone is linked to the export of the adjacent non-flow-based zone though Eq. \ref{eq:modfb:ex1b}, essentially the net position of the virtual bidding zone is limited to the NTC-values.
        \item \textbf{PTDF-based net position effect on CNECs}: The net positions of the \emph{physical and virtual} flow-based zones (zone 1-3 and VBZ 1-3, see Fig. \ref{fig:grid}) are limited by their PTDF-based effects on the CNECs' RAMs, see Eq. \ref{eq:modfb:ramposahc} and \ref{eq:modfb:ramnegahc}
        \item \textbf{Allocation constraint / external constraints}: The NTC-coupled border may be governed by a different capacity calculation region. Hence, setting the NTC-values may be handled by different processes. Importantly, the NTC limitation may only become visible to the flow-based capacity calculation region \emph{during} market coupling. If an additional constraint to the net position of VBZs is to be added - for reasons of ex-ante visibility or others - this can be done by an allocation constraint or external constraint. This is not done in this study, but (further) lower and upper bounds for the VBZ's net position can be provided to the model.
    \end{itemize}
\end{itemize}

\subsection{Other model assumptions}
\label{setup:sub5}

The following list of model assumptions concludes the setup description

\begin{itemize}[itemsep=0.1em, parsep=0pt, label={--}]
    \item \textbf{GSK}: The model uses a "flat" strategy, weighting all nodes with conventional power plants equally in each zone.
    \item \textbf{NTCs}: Throughout, the NTCs for the border of "Import/Export 1" are set to 1750 MW (ca. equal to one of the connecting CNECs of VBZ 1) and for the other borders to 2500 MW.
    \item \textbf{Congestion management costs} (penalty terms in objective function):
    \begin{itemize}[itemsep=0.1em, parsep=0pt, label={--}]
    \item Pos. RD with units in FB zones: $c^{\mathrm{RD, pos}}_{t,p} = 100 + 1.2 \cdot c^{\mathrm{var}}_{t,p}$ (for non-FB units: $500 + ...$)
    \item Neg. RD with units in FB zones: $c^{\mathrm{RD, neg}}_{t,p} = 100 + max(1.2 \cdot c^{\mathrm{var}}_{t}) - 1.2 \cdot c^{\mathrm{var}}_{t,p}$ (for non-FB units: $500 + ...$)
    \item Curtailment (after market coupling): $c^{\mathrm{curt}} = 1500$
    \item Curtailment (for D2CF and during market coupling): $c^{\mathrm{curt}} = 0$
\end{itemize}
\end{itemize}

During congestion management, the variable costs for redispatching (RD) units are multiplied by 1.2 to account for the fact that units' remuneration often exceeds their market bidding costs. The positive penalty costs for both, positive and negative RD, prevent the model from optimizing market results. High penalties for curtailing renewable energy sources after market coupling reflects the preference to handle congestions through redispatch first and only use curtailment if its effect greatly surpasses the effect of negative redispatch.\footnote{Since $c^{\mathrm{curt}} = 1000$ and $c^{\mathrm{RD, neg}}_{t,p} = 100 + max(1.2 \cdot c^{\mathrm{var}}_{t}) - 1.2 \cdot c^{\mathrm{var}}_{t,p}$, the congestion-relieving effect of curtailment has to be ca. 5 to 10 times as high for the model to choose it as a remedial action.} 

To compute costs resulting from the congestion management model, for redispatch costs the variable costs are used (adjusted variable costs for pos. redispatch to create a 20\% spread between pos. and neg. redispatch), resulting in positive costs for positive redispatch and negative costs for negative redispatch. Curtailed units are remunerated by the zonal market price, derived as the shadow price of the zonal balance during market coupling. If this is negative, the remuneration is set to zero. 

\section{Results and implications}
\label{sec:results}

This section starts by presenting the resulting market coupling and congestion management costs in an aggregated way. The section then explores the results in further detail to identify explanations and reasons.

\begin{figure}[h!]
\includegraphics[width=0.95\textwidth]{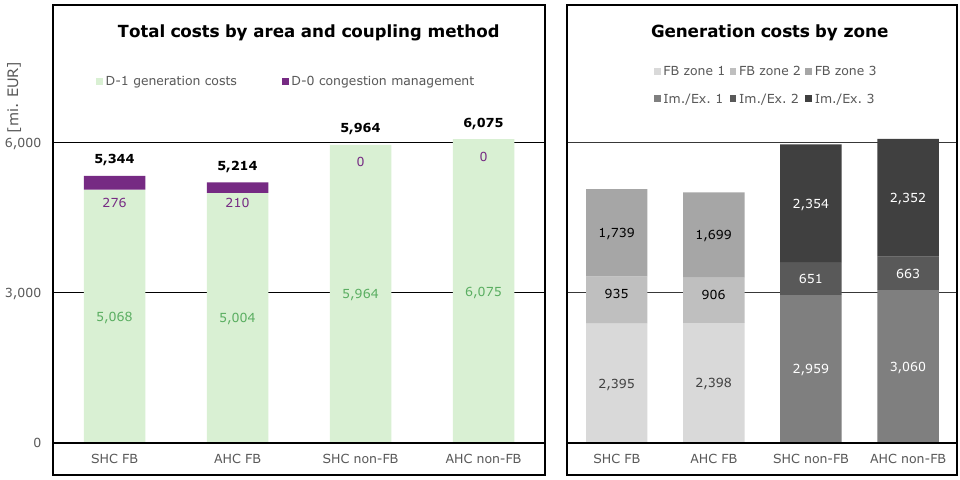}
\centering
\caption{Aggregated results of the study. \textit{Left}: Generation costs after day-ahead market coupling and congestion management cost, separately for flow-based/non-flow-based areas (aggregated) and AHC/SHC. The percentage increase/decrease from SHC to AHC is indicated. \textit{Right}: Generation costs after day-ahead market coupling, separately for each zone and AHC/SHC}
\label{fig:results}
\end{figure}

The left side of Fig. \ref{fig:results} shows the changes in D-1 generation costs and D-0 congestion management costs when switching from SHC to AHC. The results are separated between the FB region and the non-FB region. All congestion management costs are attributed to the flow-based region. Congestions occur solely within the FB region, thus all remedial actions are activated to resolve congestions within the FB region. The right side of Fig. \ref{fig:results} details the generation costs by zone, separately for FB and non-FB zones. 

Fig. \ref{fig:results} highlights that the introduction of AHC leads to a reduction in generation costs \emph{and} congestion management costs within the flow-based region but increases the generation costs in the adjacent non-FB zones. In other words, \emph{AHC reduces the dispatch costs in the flow-based region, leading to welfare gains}. Also, \emph{AHC better represents congestions within the FB region, which leads to a decrease in congestion management costs}. It becomes evident that the cost effects are distributed unequally, i.e. zones are subjected to an increase or decrease in generation costs through AHC. However, the increase in generation costs in FB zone 1 and decrease in generation costs in non-FB zone "Import/Export 3" are minimal and generally generation costs decrease in FB zones and increase in non-FB zones.

Please note that the model has several limitations, which have to be kept in mind when interpreting the results. Importantly, there are only three zone within the FB CCR and the non-FB zones can only trade with or across the FB CCR. Any limitation on the NTC-coupled border of a non-FB zone has an immediate effect on the zone's market coupling results, with no balanced-out effects on other borders of this zone. This effect is most notable for zone "Import/Export 1", which is subjected to the largest increase in generation costs. Note that in this study, the NTCs are the same in both SHC and AHC setup. On some European borders, NTCs could increase with the introduction of AHC because with AHC, NTCs may cease to consider FB CCR grid limitations, as done on some borders in the SHC setup, if VBZs are anticipated to fully capture these FB CCR limitations. This is not considered in this study and could therefore explain the increased generation costs in the non-FB zones.

Also, the results depend on assumptions:
\begin{itemize}[itemsep=0.1em, parsep=0pt, label={--}]
\item \textbf{D-2}: The D2CF is subject to several uncertainties in reality. In this analysis, they are mostly represented by uncertain renewable energy feed-in. The D2CF uncertainty, captured in part by $f_{t,l}^{\mathrm{uaf}}$, is likely underestimated in this analysis. The effect of AHC could therefore be greater in reality, as AHC removes the uncertainty introduced by $f_{t,l}^{\mathrm{uaf}}$.
\item \textbf{D-1}: The market coupling results depend on NTCs as well as flow-based parameters, most notably GSKs and the combination of CNEs and contingencies. Changing these parameters would affect the results.
\item \textbf{D-0}: The utilization of redispatch and curtailment is determined  through the chosen penalty costs in the model. In reality, other processes and restrictions apply.
\end{itemize}

\begin{figure}[h!]
\includegraphics[width=0.95\textwidth]{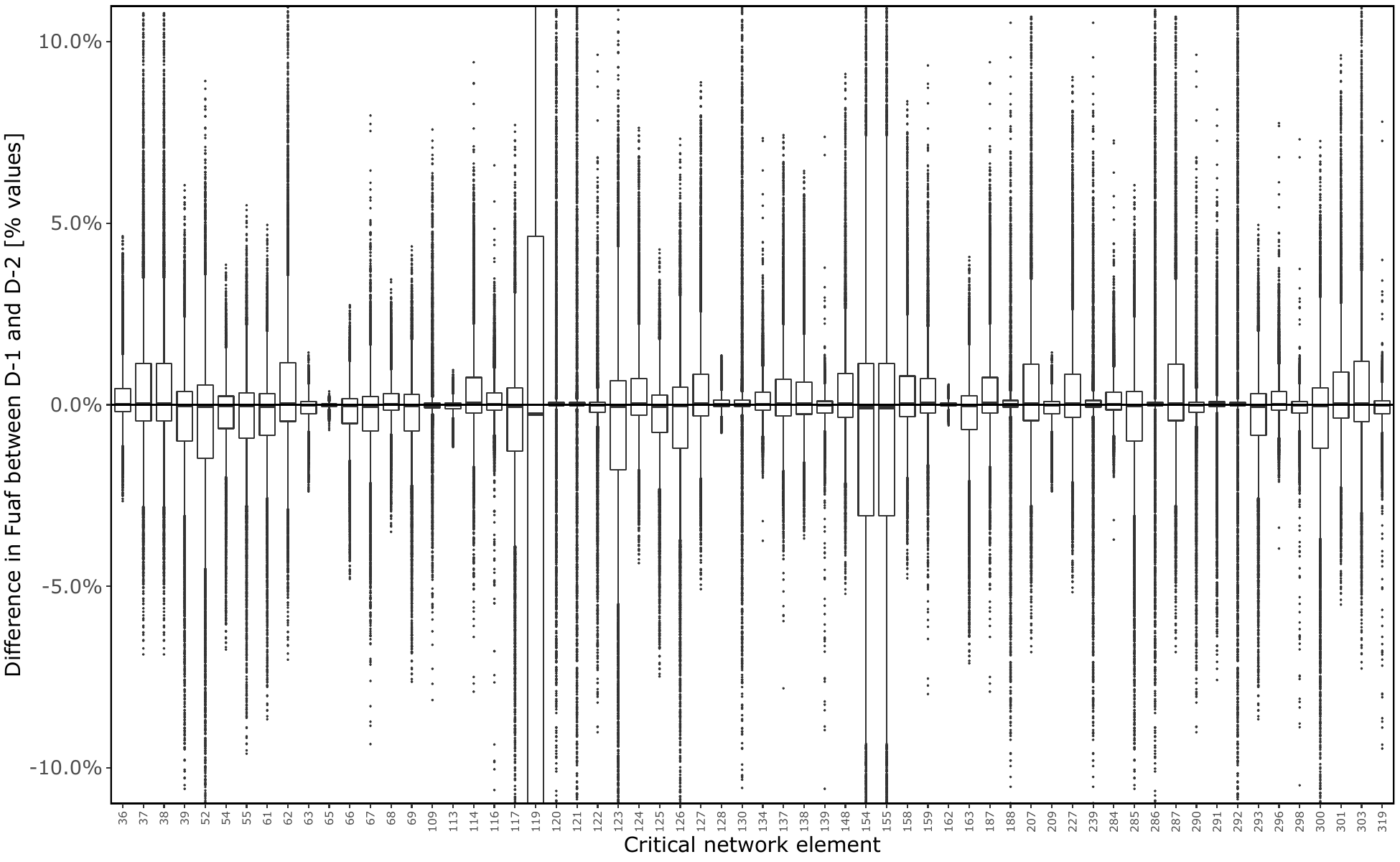}
\centering
\caption{Difference between $f_{t,l}^{\mathrm{uaf}}$ as a D2CF and a SHC market coupling result, depicted for all CNEs considered in SHC. Each boxplot shows the distribution of differences for the 8760 analyzed hours.}
\label{fig:diff_fuaf}
\end{figure}

Part of the explanation for the main results is already mentioned above, namely the presence (absence) of $f_{t,l}^{\mathrm{uaf}}$ in SHC (AHC). Fig. \ref{fig:diff_fuaf} shows the difference between the $f_{t,l}^{\mathrm{uaf}}$ as a result of market coupling and as an assumption within D2CF.\footnote{Note that $f_{t,l}^{\mathrm{uaf}}$ is typically an input for market coupling so a post-D2CF computation, not a post-market coupling computation. For this evaluation, we use $f_{t,l}^{\mathrm{uaf}}$ for both D-2 and D-1 to depict the effect of exchanges across NTC-coupled borders, as forecasted in D-2 and as a result after market coupling.} It becomes evident that $f_{t,l}^{\mathrm{uaf}}$ is subject to different degrees of uncertainty for different CNEs. The ouliers exceed +/- 5\% for almost all CNEs. The whiskers of a substantial subset of CNEs surpass +/- 2\%. This gives an indication of how uncertain $f_{t,l}^{\mathrm{uaf}}$ is. CNEs with large boxes, e.g. line 119, 154, 155, are in close proximity or directly connected to interconnector nodes on the flow-based side of NTC-coupled borders (VBZs), which explains the large uncertainties due to $f_{t,l}^{\mathrm{uaf}}$. If the D-2 forecast for the exchange across NTC-coupled borders is deviating substantially from the actual trade during D-1 market coupling, this is reflected in a large $f_{t,l}^{\mathrm{uaf}}$-deviation. 

This highlights the inefficiency in the SHC setup. Both errors introduce inefficiency into the FB setup. If $f_{t,l}^{\mathrm{uaf}}$ is \emph{overestimated} in the same burdening direction as the effect of flow-based trade, it unnecessarily reduces the possibility of additional trade between flow-based zones. If $f_{t,l}^{\mathrm{uaf}}$ is \emph{underestimated} in the same burdening direction as the effect of flow-based trade (or even estimated in the wrong direction of burden), the actual effects of trade across NTC-coupled borders due to market coupling results are greater and have to be compensated by (costly) remedial actions.

The presence of $f_{t,l}^{\mathrm{uaf}}$ in SHC, and its omission in AHC, alters the domains. Concretely, it changes the RAM given to trade activities within a flow-based region. Exemplary domains for SHC/AHC and is shown in Fig. \ref{fig:domains}. The comparison shows that the AHC domain differs from the SHC domain due to 1) more CNECs as well as 2) altered RAMs. In summary, in some parts of the domain AHC renders possible exchanges between FB zones that exceed exchanges between FB zones in the SHC setting. Other parts of the domain are more restricted in AHC and therefore allow smaller exchanges between FB zones.

\begin{figure}[h!]
\includegraphics[width=0.95\textwidth]{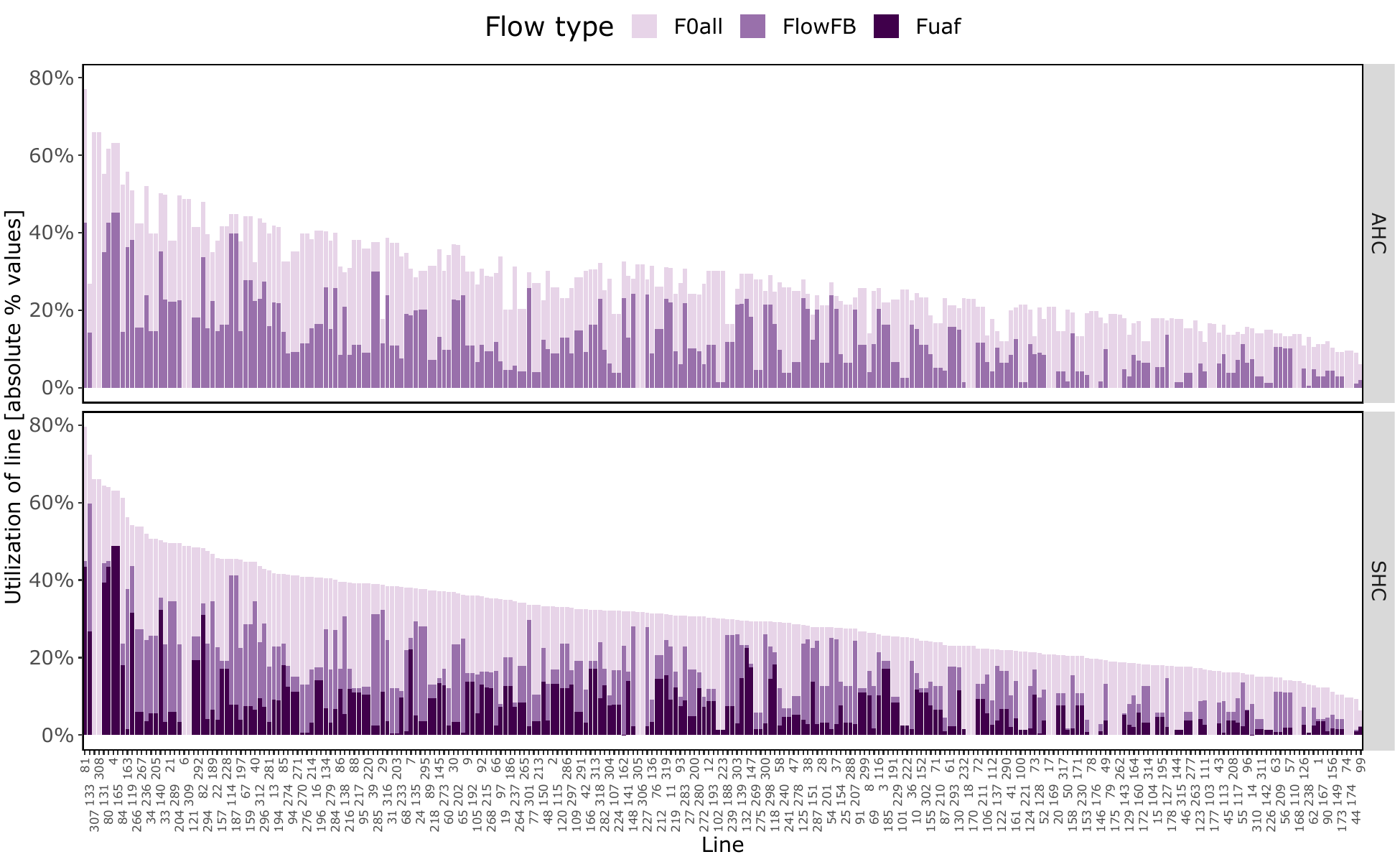}
\centering
\caption{Depicted are the average absolute values for $f_{t,l}^{\mathrm{uaf}}$, $f_{t,l}^{\mathrm{0, all}}$ and "FlowFB", the latter term of Eq. \ref{eq:f0fb}, meaning the line utilization due to trade between FB zones. The lines are sorted in descending order of the sum of the three types of utilization in the SHC setting.}
\label{fig:flows_by_run}
\end{figure}

A further decomposition of flows is shown in Fig. \ref{fig:flows_by_run}. The lower half of the figure shows that $f_{t,l}^{\mathrm{uaf}}$ amounts to significant levels of utilization for some lines. While a large share are rather unaffected by UAFs, there are several lines that reach average $f_{t,l}^{\mathrm{uaf}}$-values of over 10\%.

\begin{figure}[h!]
\includegraphics[width=0.95\textwidth]{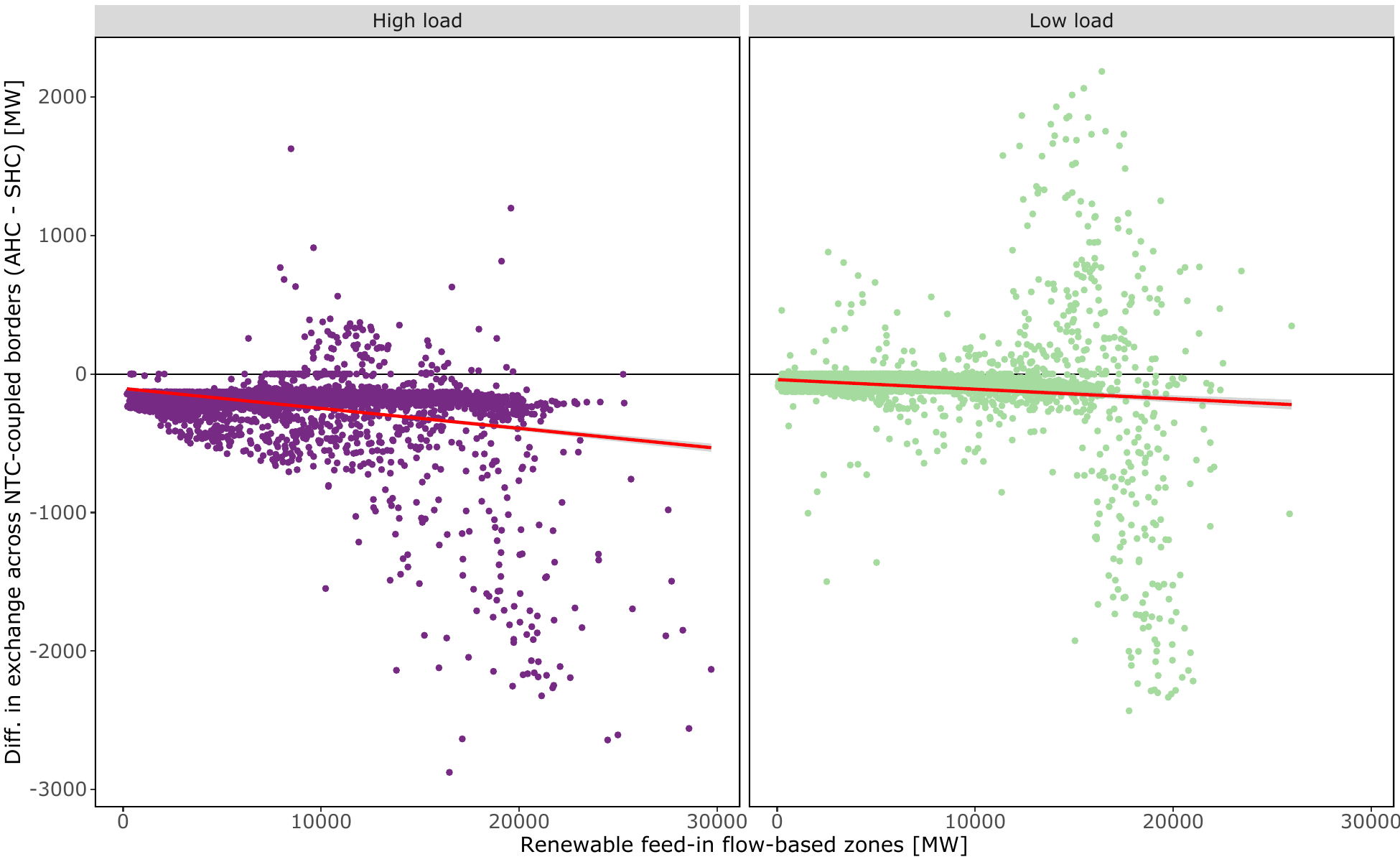}
\centering
\caption{Hourly difference in absolute exchanges across NTC-coupled borders in AHC compared to SHC (y-axis) plotted against the hourly aggregated renewable feed-in in the flow-based zones (x-axis). Positive y-values mean more exchange in AHC compared to SHC in the same hour. The observations are divided into subsets of high (above average) load on the left and low (below average) load on the right.}
\label{fig:diff_ex}
\end{figure}

To further explore, how AHC changes the trade activities, Fig. \ref{fig:diff_ex} depicts in an hourly resolution, how the exchanges across NTC-coupled borders change when moving from SHC to AHC. The difference in exchange is plotted against the aggregated renewable feed-in in the FB zones. Most points are in the negative spectrum, indicating the decrease in trade with the non-FB zones. There are, however, situations with more trade, made possible through AHC. This corresponds closely with the altered domains in Fig. \ref{fig:domains}.

As stated above, AHC maps the effect of VBZs on CNECs in the flow-based region. It also increases the amount of CNECs. While the $f_{t,l}^{\mathrm{uaf}}$ simply reserves a share of each CNEC, the trade of NTC-coupled borders is now subjected to more constraints. However, the setup of AHC and VBZs also provides the possibility of allowing more trade with across NTC-coupled borders, since the inclusion of VBZs in the market coupling optimization can find solutions where it is advantageous (welfare increasing) to trade more with non-FB zones. Concretely, \emph{trade across NTC-coupled borders is more constrained in AHC but a more accurate depiction of its effect on the FB region can lead to greater trade allocations on NTC borders}. Overall, in this study the trade between FB and non-FB zones decreases. The main beneficiaries are FB zone 2 and 3, while zone 1 sees a very slight (negligible) increase in generation costs during market coupling (cf. Fig. \ref{fig:results}). 

Fig. \ref{fig:diff_ex} also depicts a negative correlation between RES feed-in and changes in trade across NTC-coupled borders. This means that in situation of a stressed grid (high renewable feed-in) within the flow-based region, the lowering effect of AHC on trade across NTC-coupled border is the strongest. This effect appears to be exacerbated in situations of high load. In cases of low load within the FB region, AHC can result in increased trade across NTC-coupled border during more hours than in cases of high load within the FB region.

\begin{figure}[h!]
\includegraphics[width=0.95\textwidth]{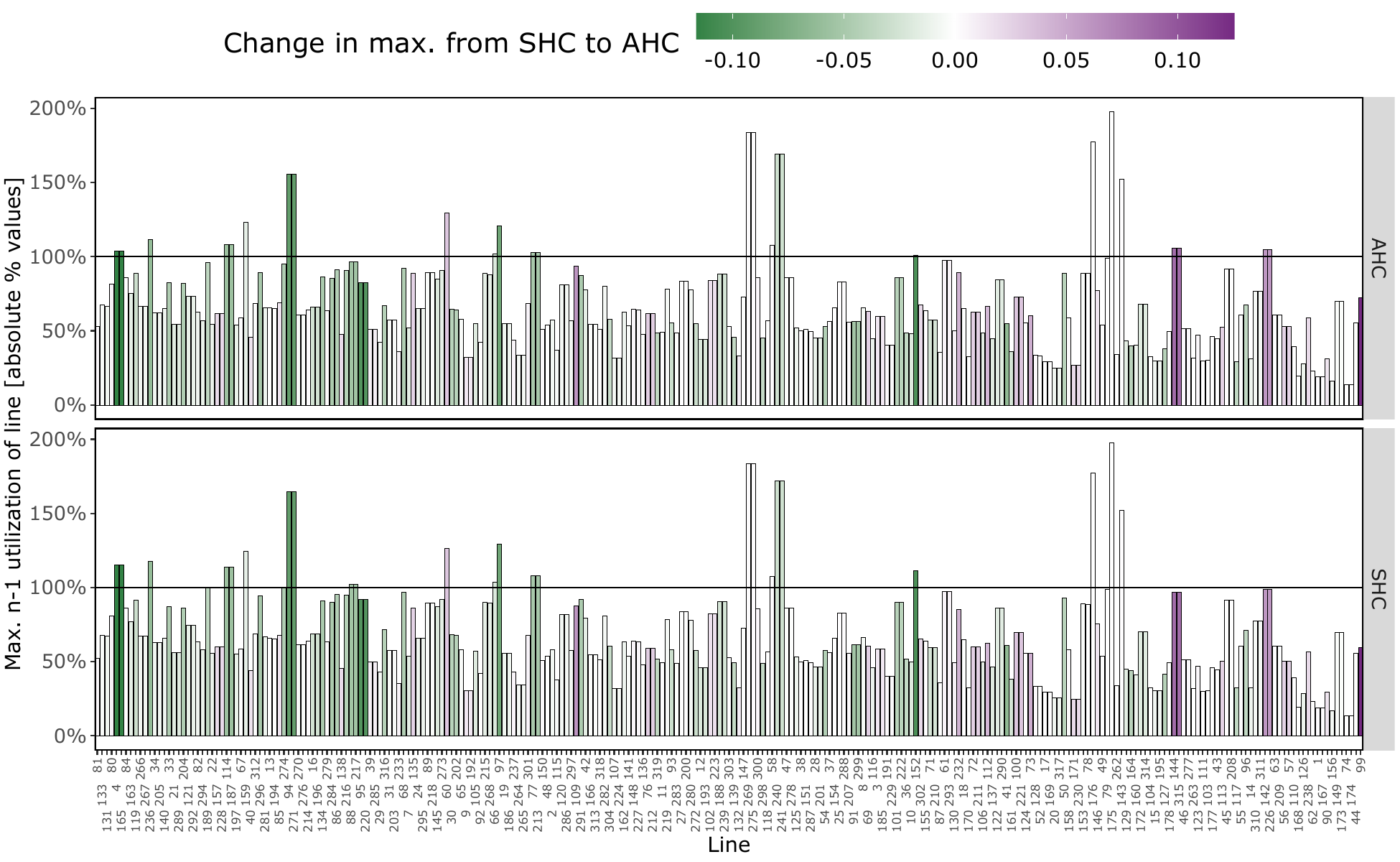}
\centering
\caption{Maximum absolute flow resulting from market coupling for AHC vs SHC. The color scale indicated the change in maxima from SHC to AHC. Negative/green values mean the line was at its maximum more burdened during SHC than AHC. The reverse is true for positive/purple values.}
\label{fig:max_flows}
\end{figure}

Not only the generation costs but also the congestion management costs decrease in the FB region through AHC. Fig. \ref{fig:max_flows} shows how the maxima in n-1 line utilization (uncorrected after market coupling and before congestion management) change between SHC and AHC. Maxima close or beyond 100\% become less common with AHC, which explains the decreasing congestion management costs with AHC. AHC leads to a better representation of the effect of trade from NTC-coupled borders, necessitating fewer remedial actions to correct the erroneous $f_{t,l}^{\mathrm{uaf}}$ (see Fig. \ref{fig:diff_fuaf}).

\section{Conclusions}
\label{sec:concl}

The advanced hybrid coupling (AHC) method is expected to enhance capacity calculation and allocation by increasing their efficiency, fairness and transparency. These gains are achieved through explicit modelling of interdependencies between cross-zonal capacities in the given flow-based capacity calculation region (CCR) and the cross-zonal capacities on its borders to adjacent CCRs. Interdependencies are modeled by including exchanges to adjacent CCRs into the flow-based method of the given CCR, using the concept of virtual bidding zones.
In this way it is possible to exploit all advantages of flow-based method, which were described in Sections \ref{sec:intro} and \ref{sec:shc_ahc}:
\begin{itemize}[itemsep=0.1em, parsep=0pt, label={--}]
    \item \textbf{Fairness gains} stem from \emph{providing the full capacity of critical network elements} with contingencies (CNECs) to all market participants (e.g. no splitting or "ex-ante" reservation), which then compete for scarce capacities in the market based on their bids.
    \item \textbf{Efficiency gains} stem from \emph{eliminating the need for capacity splitting/reservation}, in the form of unscheduled allocated flows, which is prone to forecast errors. This leads to optimal allocation and increase in overall welfare.
    \item \textbf{Transparency gains} stem from \emph{the more detailed modeling of physical power flows} in the flow-based method, enabling the identification of congested grid elements limiting cross-zonal trade.
\end{itemize}

Section \ref{sec:shc_ahc} also provides a detailed description on how different parameters and steps in the capacity calculation and allocation are impacted by AHC,
and how this contributes to the mentioned efficiency and transparency gains. AHC mainly affects the CNEC selection, the remaining available margins and the allocation constraints. Overall, AHC changes the flow-based domains and will increase the complexity of the market coupling algorithm though the introduction of virtual bidding zones. 

Section \ref{sec:setup} introduces a model, used to exemplary assess the mentioned efficiency gains in Section \ref{sec:results}, as well as to depict the previously described impact on capacity calculation parameters. The model confirms the mentioned hypothesis and shows that in the flow-based CCR the generation costs during market coupling and the congestion management costs decrease with the introduction of AHC. This leads to welfare gains. In the zones coupled by NTCs (net transfer capacities) outside the flow-based CCR, generation costs increase. It is important to recap that the model has limitations, which could partially or fully explain the increase in generation costs in these zones. This is not assumed to necessarily be a universally true result.
Overall costs, namely generation costs and congestion management costs across all zones, decrease with AHC, showing that it is the more efficient hybrid coupling method. The theoretically explored explanation is actualized in the results, i.e. a better representation of trade across NTC-coupled borders by means of virtual bidding zones and their net positions' PTDF-based (power transfer distribution factor) effect on CNECs in the flow-based CCR. This removes the inefficiency and uncertainty of unscheduled allocated flows. Thereby, AHC also allows for a better representation of congestions within the FB CCR, which reduces the congestion management costs.

AHC will shape the European power markets over the coming years. AHC is not a single "on switch", but it will rather be implemented in phases. NTC-coupled borders will introduce AHC at different times. Even the two sides of some NTC borders will introduce AHC at different times, creating times of "one-sided" AHC. Thus, the effect of AHC will unfold in several steps. Each one is expected to be a step toward a more efficient market coupling.
\newline
\newline

\newpage

\bibliographystyle{apalike}

\begin{spacing}{1}
\footnotesize
\bibliography{biblio}

\begin{thebibliography}{}

\bibitem[ACER, 2016]{acer_2016}
ACER (2016).
\newblock {Market Monitoring Report 2016}.
\newblock \url{https://acer.europa.eu/Official_documents/Acts_of_the_Agency/Publication/ACER%20Market%20Monitoring%20Report%202016%20-%20ELECTRICITY.pdf}, accessed on 20.12.2023.

\bibitem[ACER, 2019]{entsoe_1}
ACER (2019).
\newblock {ACER Decision on Core CCM: Annex I -- Day-ahead capacity calculation methodology of the Core capacity calculation region}.
\newblock \url{https://eepublicdownloads.entsoe.eu/clean-documents/nc-tasks/CORE%20-%20ANNEX%20I_III.pdf}, accessed on 04.11.2023.

\bibitem[ACER, 2023]{acer_action_plans}
ACER (2023).
\newblock {Overview of Action Plans and Derogations for the period 2020-2023}.
\newblock \url{https://acer.europa.eu/sites/default/files/documents/Official_documents/Acts_of_the_Agency/Publications%20Annexes/ACER%20Report%20on%20the%20result%20of%20monitoring%20the%20MACZT%20Generic/ACER%20Report%20on%20the%20result%20of%20monitoring%20the%20MACZT%20Derogations.pdf}, accessed on 27.02.2024.

\bibitem[{Core TSOs}, 2021]{core_fbmc_workshop}
{Core TSOs} (2021).
\newblock {Introduction to Core FB MC - Workshop}.
\newblock \url{https://hupx.hu/uploads/h%C3%ADrek/Introduction%20to%20Core%20FBMC%20workshop%20-%2022%20Nov%202021.pdf}, accessed on 27.02.2024.

\bibitem[{Core TSOs}, 2022]{entsoe_2}
{Core TSOs} (2022).
\newblock {Explanatory document to the second amendment of the Day-Ahead Capacity Calculation Methodology of the Core Capacity Calculation Region}.
\newblock \url{https://consultations.entsoe.eu/markets/core_2nd_dafb_ccm/supporting_documents/Explanatory%20Document_2nd_amendment_Core%20DA%20CCM%20.pdf}, accessed on 09.11.2023.

\bibitem[{Core TSOs}, 2023]{entsoe_3}
{Core TSOs} (2023).
\newblock {Second amendment of the Day-Ahead Capacity Calculation Methodology of the Core Capacity Calculation Region}.
\newblock \url{https://www.acm.nl/system/files/documents/bijlage-2-voorstel-tweede-wijziging-day-ahead-capaciteitsberekening-core.pdf}, accessed on 27.02.2024.

\bibitem[{Council of the European Union} and {European Parliament}, 2019]{union2019regulation}
{Council of the European Union} and {European Parliament} (2019).
\newblock {Regulation (EU) 2019/943 of the European Parliament and of the Council of 5 June 2019 on the internal market for electricity}.
\newblock {\em Official Journal of the European Union L}, 158:54--124.
\newblock Accessed on 05.10.2020.

\bibitem[{CWE TSOs}, 2011]{cwe_enhanced_fbmc_report_2011}
{CWE TSOs} (2011).
\newblock {CWE Enhanced Flow-Based MC feasibility report v2.0}.
\newblock \url{https://www.amprion.net/Dokumente/Strommarkt/Engpassmanagement/CWE-Market-Coupling/CWE_FB-MC_feasibility_report_2.0_19102011.pdf}, accessed on 20.12.2023.

\bibitem[{CWE TSOs}, 2014]{cwe_fbmc_approval_doc_2014}
{CWE TSOs} (2014).
\newblock {CWE FBMC approval document v2.1}.
\newblock \url{https://www.bundesnetzagentur.de/SharedDocs/Downloads/DE/Sachgebiete/Energie/Unternehmen_Institutionen/NetzzugangUndMesswesen/Marktkopplung/140530%20%20CWE%20FB%20MC%20Approval%20document.pdf?__blob=publicationFile&v=1}, accessed on 20.12.2023.

\bibitem[{Ember}, 2023]{ember}
{Ember} (2023).
\newblock {Electricity Interconnection in Europe}.
\newblock \url{https://ember-climate.org/data/data-tools/electricity-interconnection-europe/}, accessed on 22.12.2023.

\bibitem[{ENTSO-E}, 2022]{ccrs}
{ENTSO-E} (2022).
\newblock {Determination of capacity calculation regions: All TSOs’ proposal for amendment of the Determination of capacity calculation regions methodology in accordance with Article 15(1) of the Commission Regulation (EU) 2015/1222 of 24 July 2015 establishing a guideline on capacity allocation and congestion management (12 October 2022)}.
\newblock \url{https://www.nve.no/media/14782/proposal-from-statnett-to-rme-on-determination-of-ccrs.pdf}, accessed on 21.12.2023.

\bibitem[{ENTSO-E}, 2023]{tyndp}
{ENTSO-E} (2023).
\newblock {TYNDP 2022 Projects Sheets}.
\newblock \url{https://tyndp2022-project-platform.azurewebsites.net/projectsheets}, accessed on 21.12.2023.

\bibitem[Guo et~al., 2009]{guo2009direct}
Guo, J., Fu, Y., Li, Z., and Shahidehpour, M. (2009).
\newblock Direct calculation of line outage distribution factors.
\newblock {\em IEEE Transactions on Power Systems}, 24(3):1633--1634.

\bibitem[{MapCharts}, 2023]{MapCharts}
{MapCharts} (2023).
\newblock Europe.
\newblock \url{https://www.mapchart.net/europe.html}, accessed on 21.12.2023.

\bibitem[{Nordic CCM Project}, 2023]{nordic_golive_update}
{Nordic CCM Project} (2023).
\newblock {News Update: Nordic Flow-based Market Coupling Go-live expected October 2024}.
\newblock \url{https://nordic-rcc.net/wp-content/uploads/2023/11/News-update-7-November-2023.pdf}, accessed on 27.02.2024.

\bibitem[Sch{\"o}nheit et~al., 2021]{schonheit2021toward}
Sch{\"o}nheit, D., Kenis, M., Lorenz, L., M{\"o}st, D., Delarue, E., and Bruninx, K. (2021).
\newblock Toward a fundamental understanding of flow-based market coupling for cross-border electricity trading.
\newblock {\em Advances in Applied Energy}, 2:100027.

\end{thebibliography}
\end{spacing}

\newpage
\appendix

\section{Model documentation}
\label{app:model}

\subsection{Two-days ahead base case}
\label{subsec:mod_d-2}

{
\allowdisplaybreaks
{\small \setstretch{1.1}
\begin{subequations}\label{eq:mod_d-2}
\begin{align}
\omit\rlap{\textit{\underline{Objective}}} & \omit\rlap{\textit{Minimization of generation and curtailment costs}} \notag \\
\min \quad TC^\mathrm{D-2} & = \sum_{t,z} (CG_{t,z}^\mathrm{D-2} + CC_{t,z}^\mathrm{D-2}) & & \label{eq:modb:tc} \\
\omit\rlap{\textit{\underline{Subject to}}} & \omit\rlap{\textit{Generation costs}} \notag \\
CG_{t,z}^\mathrm{D-2} & = \sum_{p \in mp(z)} G_{t,p}^\mathrm{D-2} \cdot c_{t,p}^\mathrm{var} & \forall t \in T & \enspace \forall z \in Z \label{eq:modb:cg} \\
& \omit\rlap{\textit{Congestion costs}} \notag \\
CC_{t,z}^\mathrm{D-2} & = \sum_{n \in mn(z)} CURT_{t,n}^\mathrm{D-2} \cdot c^{\mathrm{curt}} & \forall t \in T & \enspace \forall z \in Z \label{eq:modb:cc} \\
& \omit\rlap{\textit{Limitation of curtailment to renewable feed-in}} \notag \\
CURT_{t,n}^\mathrm{D-2} & \leq ren_{t,n}^\mathrm{D-2} & \forall t \in T & \enspace \forall n \in N \label{eq:modb:curtmax} \\
& \omit\rlap{\textit{Maximum generation is the capacity}} \notag \\
G_{t,p}^\mathrm{D-2} & \leq g_{p}^\mathrm{max} & \forall t \in T & \enspace \forall p \in P \label{eq:modb:genmax} \\
& \omit\rlap{\textit{Energy balance for every FB zone}} \notag \\
\sum_{n \in mn(z)} d_{t,n} & = \sum_{n \in mn(z)} \rlap{$\Big[\sum_{p \in mp(n)} G_{t,p}^\mathrm{D-2} + ren_{t,n}^\mathrm{D-2} - CURT_{t,n}^\mathrm{D-2} \Big]$} \notag \\
& \quad + \rlap{$\sum_{x \in mz(z)} EX_{t,x}^\mathrm{D-2} - NP_{t,z}^\mathrm{D-2}$} & \forall t \in T & \enspace \forall z \in Z^{\mathrm{FB}} \label{eq:modb:bal} \\ 
& \omit\rlap{\textit{Energy balance for every non-FB zone}} \notag \\
\sum_{n \in mn(z)} d_{t,n} & = \sum_{n \in mn(z)} \rlap{$\Big[\sum_{p \in mp(n)} G_{t,p}^\mathrm{D-2} + ren_{t,n}^\mathrm{D-2} - CURT_{t,n}^\mathrm{D-2} \Big]$} \notag \\
& \quad - \rlap{$EX_{t,z}^\mathrm{D-2}$} & \forall t \in T & \enspace \forall z \not\in Z^{\mathrm{FB}} \label{eq:modb:bal} \\ 
& \omit\rlap{\textit{Export balance (non-FBMC zones) limited to day-ahead NTC-values}} \notag \\
EX_{t,z}^\mathrm{D-2} & \leq ntc_{t,z}  & \forall t \in T & \enspace z \not\in Z^{\mathrm{FB}} \label{eq:modb:ex2} \\
EX_{t,z}^\mathrm{D-2} & \geq -ntc_{t,z}  & \forall t \in T & \enspace z \not\in Z^{\mathrm{FB}} \label{eq:modb:ex3} \\
& \omit\rlap{\textit{Sum of net positions within flow-based area is zero}} \notag \\
\sum_{z \in Z^{\mathrm{FB}}} NP_{t,z}^\mathrm{D-2} & = 0 & \enspace \forall t \in T & \label{eq:modb:np0a} 
\end{align}
\end{subequations}
}}

\newpage
\subsection{Day-ahead market coupling: Standard hybrid coupling}
\label{subsec:mod_d-1shc}

{
\allowdisplaybreaks
{\small \setstretch{1.1}
\begin{subequations}\label{eq:modm}
\begin{align}
\omit\rlap{\textit{\underline{Objective}}} & \omit\rlap{\textit{Minimization of generation and congestion costs}} \notag \\
\min \quad TC^\mathrm{D-1} & = \sum_{t,z} (CG_{t,z}^\mathrm{D-1} + CC_{t,z}^\mathrm{D-1}) & & \label{eq:modm:tc} \\
\omit\rlap{\textit{\underline{Subject to}}} & \omit\rlap{\textit{Generation costs}} \notag \\
CG_{t,z}^\mathrm{D-1} & = \sum_{p \in mp(z)} G_{t,p}^\mathrm{D-1} \cdot c_{t,p}^\mathrm{var} & \forall t \in T & \enspace \forall z \in Z \label{eq:modm:cg} \\
& \omit\rlap{\textit{Congestion costs}} \notag \\
CC_{t,z}^\mathrm{D-1} & = \sum_{n \in mn(z)} CURT_{t,n}^\mathrm{D-1} \cdot c^{\mathrm{curt}} & \forall t \in T & \enspace \forall z \in Z \label{eq:modm:cc} \\
& \omit\rlap{\textit{Limitation of curtailment to renewable feed-in}} \notag \\
CURT_{t,n}^\mathrm{D-1} & \leq ren_{t,n} & \forall t \in T & \enspace \forall n \in N \label{eq:modm:curt} \\
& \omit\rlap{\textit{Maximum generation is the capacity}} \notag \\
G_{t,p}^\mathrm{D-1} & \leq g_{p}^\mathrm{max}  & \forall t \in T & \enspace \forall p \in P \label{eq:modm:genmax} \\
& \omit\rlap{\textit{Energy balance for every FB zone}} \notag \\
\sum_{n \in mn(z)} d_{t,n} & = \sum_{n \in mn(z)} \rlap{$\Big[\sum_{p \in mp(n)} G_{t,p}^\mathrm{D-1} + ren_{t,n} - CURT_{t,n}^\mathrm{D-1} \Big]$} \notag \\
& \quad + \rlap{$\sum_{x \in mz(z)} EX_{t,x}^\mathrm{D-1} - NP_{t,z}^\mathrm{D-1}$} & \forall t \in T & \enspace \forall z \in Z^{\mathrm{FB}} \label{eq:modfb:bal} \\ 
& \omit\rlap{\textit{Energy balance for every non-FB zone}} \notag \\
\sum_{n \in mn(z)} d_{t,n} & = \sum_{n \in mn(z)} \rlap{$\Big[\sum_{p \in mp(n)} G_{t,p}^\mathrm{D-1} + ren_{t,n} - CURT_{t,n}^\mathrm{D-1} \Big]$} \notag \\
& \quad - \rlap{$EX_{t,z}^\mathrm{D-1}$} & \forall t \in T & \enspace \forall z \not\in Z^{\mathrm{FB}} \label{eq:modfb:bal} \\ 
& \omit\rlap{\textit{Export balance (non-FBMC zones) limited to day-ahead NTC-values}} \notag \\
EX_{t,z}^\mathrm{D-1} & \leq ntc_{t,z}  & \forall t \in T & \enspace z \not\in Z^{\mathrm{FB}} \label{eq:modfb:ex2} \\
EX_{t,z}^\mathrm{D-1} & \geq -ntc_{t,z}  & \forall t \in T & \enspace z \not\in Z^{\mathrm{FB}} \label{eq:modfb:ex3} \\
& \omit\rlap{\textit{Sum of net positions within flow-based area is zero}} \notag \\
\sum_{z \in Z^{\mathrm{FB}}} NP_{t,z}^\mathrm{D-1} & = 0 & \enspace \forall t \in T & \label{eq:modfb:np0a} \\
& \omit\rlap{\textit{Limitation of positive and negative flow changes on critical network elements}} \notag \\
\overline{ram}_{j,t}^{\mathrm{pos, SHC}} & \geq  \sum_{z \in Z^{\mathrm{FB}}} \Big[ptdf_{j,z}^{\mathrm{Z}} \cdot NP_{t,z}^\mathrm{D-1}\Big] & \forall t \in T & \enspace \forall j \in J \label{eq:modfb:rampos} \\
\overline{ram}_{j,t}^{\mathrm{neg, SHC}} & \leq \sum_{z \in Z^{\mathrm{FB}}} \Big[ptdf_{j,z}^{\mathrm{Z}} \cdot NP_{t,z}^\mathrm{D-1}\Big] & \forall t \in T & \enspace \forall j \in J \label{eq:modfb:ramneg} 
\end{align}
\end{subequations}
}}

\newpage
\subsection{Day-ahead market coupling: Advanced hybrid coupling}
\label{subsec:mod_d-1ahc}

{
\allowdisplaybreaks
{\small \setstretch{1.1}
\begin{subequations}\label{eq:modm}
\begin{align}
\omit\rlap{\textit{\underline{Objective}}} & \omit\rlap{\textit{Minimization of generation and congestion costs}} \notag \\
\min \quad TC^\mathrm{D-1} & = \sum_{t,z} (CG_{t,z}^\mathrm{D-1} + CC_{t,z}^\mathrm{D-1}) & & \label{eq:modm:tc} \\
\omit\rlap{\textit{\underline{Subject to}}} & \omit\rlap{\textit{Generation costs}} \notag \\
CG_{t,z}^\mathrm{D-1} & = \sum_{p \in mp(z)} G_{t,p}^\mathrm{D-1} \cdot c_{t,p}^\mathrm{var} & \forall t \in T & \enspace \forall z \in Z \label{eq:modm:cg} \\
& \omit\rlap{\textit{Congestion costs}} \notag \\
CC_{t,z}^\mathrm{D-1} & = \sum_{n \in mn(z)} CURT_{t,n}^\mathrm{D-1} \cdot c^{\mathrm{curt}} & \forall t \in T & \enspace \forall z \in Z \label{eq:modm:cc} \\
& \omit\rlap{\textit{Limitation of curtailment to renewable feed-in}} \notag \\
CURT_{t,n}^\mathrm{D-1} & \leq ren_{t,n} & \forall t \in T & \enspace \forall n \in N \label{eq:modm:curt} \\
& \omit\rlap{\textit{Maximum generation is the capacity}} \notag \\
G_{t,p}^\mathrm{D-1} & \leq g_{p}^\mathrm{max}  & \forall t \in T & \enspace \forall p \in P \label{eq:modm:genmax} \\
& \omit\rlap{\textit{Energy balance for every physical FB zone}} \notag \\
\sum_{n \in mn(z)} d_{t,n} & = \sum_{n \in mn(z)} \rlap{$\Big[\sum_{p \in mp(n)} G_{t,p}^\mathrm{D-1} + ren_{t,n} - CURT_{t,n}^\mathrm{D-1} \Big]$} \notag \\
& \quad - \rlap{$NP_{t,z}^\mathrm{D-1}$} & \forall t \in T & \enspace \forall z \in Z_{\mathrm{AHC, phys}}^{\mathrm{FB}} \label{eq:modfb:bal} \\ 
& \omit\rlap{\textit{Energy balance for every non-FB zone}} \notag \\
\sum_{n \in mn(z)} d_{t,n} & = \sum_{n \in mn(z)} \rlap{$\Big[\sum_{p \in mp(n)} G_{t,p}^\mathrm{D-1} + ren_{t,n} - CURT_{t,n}^\mathrm{D-1} \Big]$} \notag \\
& \quad - \rlap{$EX_{t,z}^\mathrm{D-1}$} & \forall t \in T & \enspace \forall z \not\in Z^{\mathrm{FB}} \label{eq:modfb:bal} \\ 
& \omit\rlap{\textit{Net positions for virtual bidding zones}} \notag \\
NP_{t,z}^\mathrm{D-1} & = \sum_{x \in mvbz(z)}  EX_{t,x}^\mathrm{D-1} & \forall t \in T & \enspace z \in Z_{\mathrm{AHC, virt}}^{\mathrm{FB}} \label{eq:modfb:ex1b} \\
& \omit\rlap{\textit{Export balance (non-FBMC zones) limited to day-ahead NTC-values}} \notag \\
EX_{t,z}^\mathrm{D-1} & \leq ntc_{t,z}  & \forall t \in T & \enspace z \not\in Z^{\mathrm{FB}} \label{eq:modfb:ex2ahc} \\
EX_{t,z}^\mathrm{D-1} & \geq -ntc_{t,z}  & \forall t \in T & \enspace z \not\in Z^{\mathrm{FB}} \label{eq:modfb:ex3ahc} \\
& \omit\rlap{\textit{Sum of net positions within flow-based area is zero}} \notag \\
\sum_{z \in Z^{\mathrm{FB}}} NP_{t,z}^\mathrm{D-1} & = 0 & \enspace \forall t \in T & \label{eq:modfb:np0a} \\
& \omit\rlap{\textit{Limitation of positive and negative flow changes on critical network elements}} \notag \\
\overline{ram}_{j,t}^{\mathrm{pos, AHC}} & \geq  \sum_{z \in Z_{\mathrm{AHC}}^{\mathrm{FB}}} \Big[ptdf_{j,z}^{\mathrm{Z}} \cdot NP_{t,z}^\mathrm{D-1}\Big] & \forall t \in T & \enspace \forall j \in J \label{eq:modfb:ramposahc} \\
\overline{ram}_{j,t}^{\mathrm{neg, AHC}} & \leq \sum_{z \in Z_{\mathrm{AHC}}^{\mathrm{FB}}} \Big[ptdf_{j,z}^{\mathrm{Z}} \cdot NP_{t,z}^\mathrm{D-1}\Big] & \forall t \in T & \enspace \forall j \in J \label{eq:modfb:ramnegahc} 
\end{align}
\end{subequations}
}}

\newpage
\subsection{Congestion management}
\label{subsec:mod_d-0}

{
\allowdisplaybreaks
{\small \setstretch{1.1}
\begin{subequations}\label{eq:modr}
\begin{align}
\omit\rlap{\textit{\underline{Objective}}} & \omit\rlap{\textit{Minimization of congestion costs}} \notag \\
\min \quad TC^\mathrm{D-0} & = \sum_{t,z} CC_{t,z}^\mathrm{D-0} & & \label{eq:modr:tc} \\ 
\omit\rlap{\textit{\underline{Subject to}}} & \omit\rlap{\textit{Congestion costs: Redispatch (generation) and curtailment costs}} \notag \\
CC_{t,z}^\mathrm{D-0} & = \sum_{p \in mprd(z)} \Big[RD_{t,p}^\mathrm{pos} \cdot c^{\mathrm{RD, pos}}_{t,p} + RD_{t,p}^\mathrm{neg} \cdot c^{\mathrm{RD, neg}}_{t,p} \Big] \notag \\
& \quad + \sum_{n \in mn(z)} CURT_{t,n}^\mathrm{D-0} \cdot c^{\mathrm{curt}} & \forall t \in T & \enspace \forall z \in Z \label{eq:modr:cc} \\
& \omit\rlap{\textit{Curtailment is limited to remaining renewable feed-in}} \notag \\
CURT_{t,n}^\mathrm{D-0} & \leq ren_{t,n} - curt_{t,n}^\mathrm{D-1} & \forall t \in T & \enspace \forall n \in N \label{eq:modr:curtmax} \\
& \omit\rlap{\textit{Positive redispatch is limited to remaining capacity}} \notag \\
RD_{t,p}^\mathrm{pos} & \leq g_{p}^\mathrm{max} - g_{t,p}^\mathrm{D-1} & \forall t \in T & \enspace \forall p \in P^{RD} \label{eq:modr:rdpos} \\
& \omit\rlap{\textit{Negative redispatch is limited to day-ahead generation}} \notag \\
RD_{t,p}^\mathrm{neg} & \leq g_{t,p}^\mathrm{D-1} & \forall t \in T & \enspace \forall p \in P^{RD} \label{eq:modr:rdneg} \\
& \omit\rlap{\textit{Energy balance for every node}} \notag \\
d_{t,n} + INJ_{t,n}^\mathrm{D-0} & = \rlap{$\mathlarger{\sum}_r ren_{t,n,r} - curt_{t,n}^\mathrm{D-1} - CURT_{t,n}^\mathrm{D-0}$} \notag \\ 
& \quad + \rlap{$\mathlarger{\sum}_{p \in mp(n)} g_{t,p}^\mathrm{D-1} + \mathlarger{\sum}_{p \in mprd(n)} \Big[ RD_{t,p}^\mathrm{pos} - RD_{t,p}^\mathrm{neg} \Big]$} & \forall t \in T & \enspace \forall n \in N \label{eq:modr:bal} \\
& \omit\rlap{\textit{Computation of flows}} \notag \\
F_{t,l}^\mathrm{D-0} & = \sum_{n \in N} \Big[ptdf_{l,n}^{\mathrm{N}} \cdot INJ_{t,n}^\mathrm{D-0}\Big]  & \forall t \in T & \enspace \forall l \in L^\mathrm{c} \label{eq:modr:f} \\
& \omit\rlap{\textit{Sum of nodal injections}} \notag \\
\sum_{n \in N} INJ_{t,n}^\mathrm{D-0} & = 0 & \forall t \in T \label{eq:modr:inj0} \\
& \omit\rlap{\textit{Limitation of line flows to capacities (positive and negative)}} \notag \\
F_{t,l}^\mathrm{D-0} & \leq fmax_{l} - frm_{l} & \forall t \in T & \enspace \forall l \in L^\mathrm{c}  \label{eq:modr:fmax1} \\
F_{t,l}^\mathrm{D-0} & \geq -fmax_{l} + frm_{l} & \forall t \in T & \enspace \forall l \in L^\mathrm{c}  \label{eq:modr:fmax2}
\end{align}
\end{subequations}
}}

\newpage
\subsection{Visualized domains for SHC and AHC}
\label{subsec:mod_d-0}

\begin{figure}[h!]
\includegraphics[width=0.9\textwidth]{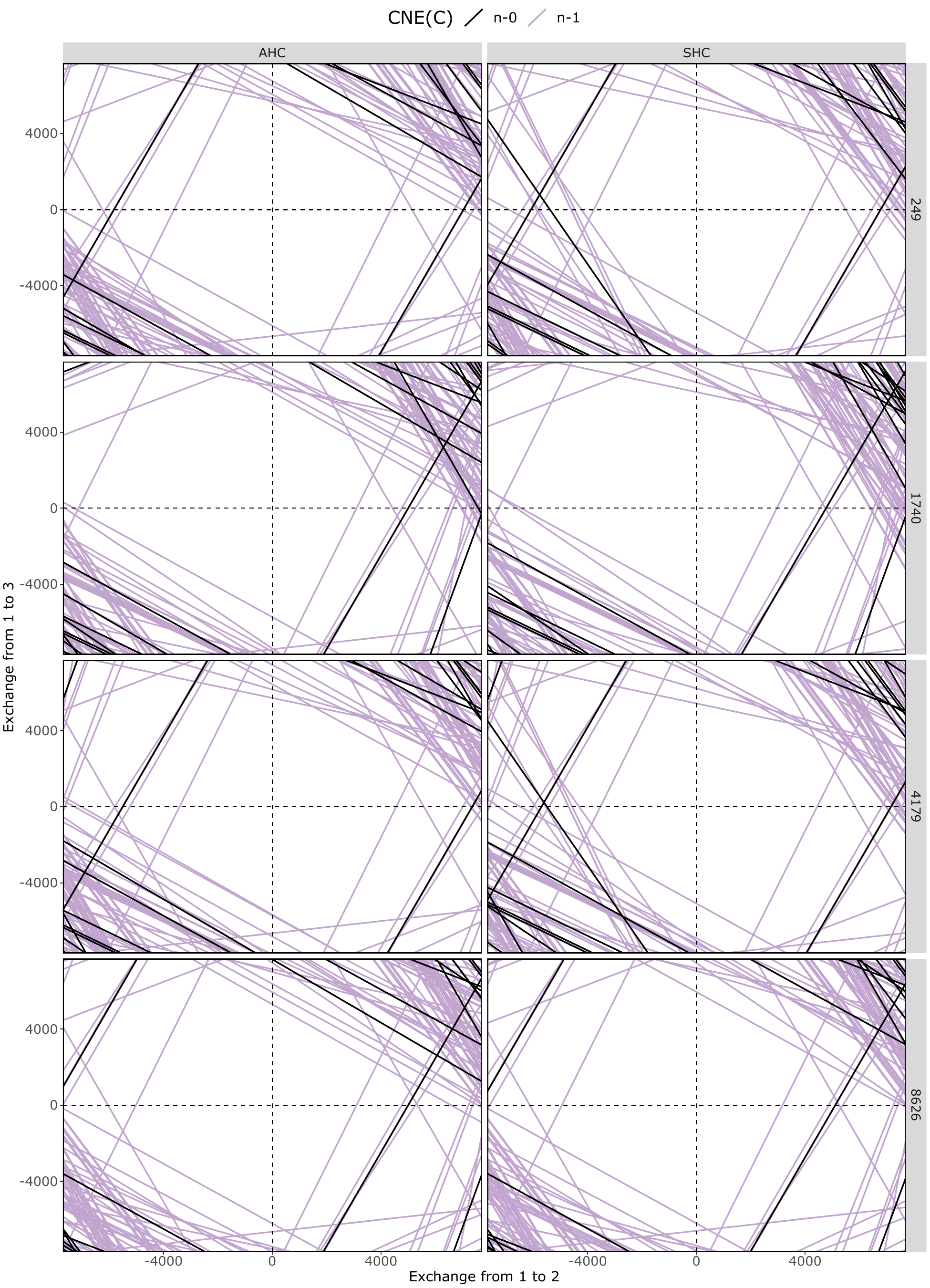}
\centering
\caption{Flow-based domains for four timestamps for AHC/SHC (left/right). The x-axis (y-axis) describes exchange from FB zone 1 to FB zone 2 (FB zone 3).}
\label{fig:domains}
\end{figure}

\end{document}